\documentclass{aa}
\usepackage{txfonts}
\usepackage{natbib}
\usepackage{graphicx}
\usepackage{amssymb}
\usepackage{lscape}
\usepackage{xspace}
\usepackage{mathrsfs}
\usepackage[dvips]{rotating}
\bibpunct{(}{)}{;}{a}{}{,}

\def\A{\r{A}\xspace}

\def\Ha{H$\alpha$\xspace}
\def\kms{$\mathrm{km}~\mathrm{s}^{-1}$\xspace}

\def\Msun{M$_\odot$\xspace}
\def\deg{$^\circ$\xspace}
\def\Av{A$_\mathrm{V}$\xspace}
\def\Rv{R$_\mathrm{V}$\xspace}
\def\OI{[O\,{\sc i}]\xspace}
\begin{document}
   \title{MWC~297: a young high-mass star rotating at critical
     velocity\thanks{Based on observations 
   made with ESO telescopes at the La Silla Paranal Observatory under
   program IDs 077.D-0071(B--C), 077.D-0095(C--F), 079.C-0012(A--H)
   and 079.C-0207(A).}}


   \author{B.~Acke
          \inst{1}\fnmsep\thanks{Postdoctoral Fellows of the Fund for
          Scientific Research, Flanders.}
          \and
	  T.~Verhoelst
          \inst{1}\fnmsep\inst{7}\fnmsep$^{\star\star}$
	  \and
	  M.E.~van~den~Ancker
	  \inst{2}
	  \and
	  P.~Deroo
	  \inst{1}
	  \and
	  C.~Waelkens
	  \inst{1}
	  \and
	  O.~Chesneau
	  \inst{3}
	  \and
	  E.~Tatulli
	  \inst{4}
	  \and
	  M.~Benisty
	  \inst{5}
	  \and
	  E.~Puga
	  \inst{1}
	  \and
	  L.B.F.M.~Waters
	  \inst{1}\fnmsep\inst{6}
	  \and
	  A.~Verhoeff
	  \inst{6}
          \and
	  A.~de~Koter
	  \inst{6}
          }

   \offprints{B.~Acke}

   \institute{Instituut voor Sterrenkunde, KULeuven,
              Celestijnenlaan 200D, B-3001 Leuven, Belgium \\
              \email{bram@ster.kuleuven.be}
         \and
             European Southern Observatory, Karl-Schwarzschild-Str. 2,
              D-85748 Garching bei M{\"u}nchen, Germany
	 \and
	     Observatoire de la C{\^o}te d'Azur/CNRS, BP 4229, 06304 Nice
             C{\'e}dex 4, France
	 \and
	     Osservatorio Astrofisico di Arcetri, L.go E. Fermi, 5,
              50125 Firenze, Italy
	 \and
	     Laboratoire d'Astrophysique de l'Observatoire de
             Grenoble, BP 53, F-38041 Grenoble C{\'e}dex 9, France
	 \and Sterrenkundig Instituut 'Anton Pannekoek', Kruislaan
              403, 1098 SJ Amsterdam, the Netherlands 
	 \and University of Manchester, Jodrell Bank Centre for
              Astrophysics, Manchester, M13 9PL, U.K.
             }

   \date{Received --; accepted --}

 
  \abstract
   {\object{MWC 297} is a nearby young massive B[e] star. The central
  star 
  is attenuated by 8 magnitudes in the optical and has a high
  projected rotational velocity of 350~km s$^{-1}$. Despite the wealth
  of published observations, the nature of this object and its
  circumstellar environment is not understood very well.}
   {With the present paper, we intend to shed light on the geometrical
  structure of the circumstellar matter that is responsible for the
  near- to mid-infrared flux excess.}
   {The H-band (1.6--2.0~$\mu$m), K-band (2.0--2.5~$\mu$m), and N-band
  (8--13~$\mu$m) brightness 
  distribution of MWC~297 was probed with the ESO interferometric
  spectrographs AMBER and MIDI, mounted on the VLTI in Paranal,
  Chile. We obtained visibility measurements on 3 AMBER and 12 MIDI
  baselines, covering a wide range of spatial frequencies. Different
  models (parametrized circumstellar disks, a dusty halo) were
  invoked 
  to fit the data, all of which fail to do so in a satisfying way. We
  approximated the brightness distribution in H, K, and N with a
  geometric model consisting of three Gaussian disks with different
  extents and brightness temperatures. This model can account for the
  entire near- to mid-IR emission of MWC~297.}
   {The circumstellar matter around MWC~297 is resolved on all
  baselines. The near- and mid-IR emission, including the silicate
  emission at 10 micron, emanates from a very compact region
  (FWHM~$<$~1.5~AU) around the central star.}
   {We argue that the extinction towards the MWC~297 star+disk system
  is interstellar and most likely due to remnants of the natal
  cloud from which MWC~297 was formed. Furthermore, we argue that the
  circumstellar matter in the MWC~297 system is 
  organized in a circumstellar disk, seen under moderate ($i <
  40^{\circ}$) inclination. The disk displays no inner emission-free
  gap at the resolution of our interferometric observations.
  The low inclination of the disk implies that the already high
  projected rotational velocity of the star corresponds to an
  actual rotational velocity that exceeds the critical velocity of the
  star. This result shows that stars can obtain such high rotation
  rates at birth. We discuss the impact of this result in terms of the
  formation of high-mass stars and the main-sequence evolution of
  classical Be stars. 
}

   \maketitle
%

\section{Introduction}

At a distance of 250~pc and with a stellar mass of approximately
10~\Msun 
\citep{drew97}, MWC~297 is one of the closest massive young stars. 
It is therefore an ideal target for studying the formation process
of massive stars. The source has
been the subject of many investigations, covering the entire spectral
range from X-rays to centimeter wavelengths \citep{vink05b, damiani06,
drew97, andrillat98, malbet07, terada01,
benedettini01,ackeiso,ackenano, henning98, manoj07, palla93, han98,
skinner93}. The image that emerges from these studies is far from
uniform.

Based on an optical spectrum, \citet{drew97} have shown convincingly
that MWC~297 is a highly attenuated B1.5V star (\Av $\approx$
8). They derive a distance to the source of 250~pc. Furthermore, these
authors indicate that the star is a rapid rotator ($\mathrm{v} \sin i
= 350 \pm 50$~\kms) and suggest that its rotation axis must be almost
in the plane 
of the sky. The MERLIN 5~GHz data, presented in the same article,
display a north-south elongation, along PA=165\deg~E of N. Drew et
al. point out that this elongation may reflect the equatorial plane of a
disk-like 
circumstellar structure around MWC~297. Both the high rotational
velocity and the elongated radio structure are consistent with the
presence of an edge-on accretion disk, which is the byproduct of the
stellar formation of MWC~297.

Near-IR interferometric observations, both broad-band
\citep{millan-gabet01, eisner03, monnier06} and spectrally dispersed
\citep{malbet07}, have shown that the radiation emitted by the hot
circumstellar dust is confined to a very small region around the
central star \citep[ring diameter of 1.4~AU,][]{monnier06}. They
suggest that the circumstellar dust around 
MWC~297 is located in an accretion disk. In a
non-accreting disk \citep[i.e. a {\em passive} disk in the terminology
of][]{dullemond01} the temperature and density structure is  
predominantly determined by the stellar irradiation. Such a disk has
a significant inner disk gap, due to dust sublimation close to
the star. The source would display a much more extended, hence
more resolved, brightness 
distribution on the sky. In contrast to the conclusion of
\citet{drew97}, \citet{malbet07} suggest an 
almost pole-on orientation of the accretion disk, based on the
analysis of one spectro-interferometric AMBER measurement and regular
spectroscopy of the H$\alpha$, H$\beta$ and Br$\gamma$ lines. Their
best model consists of a geometrically flat, optically thick accretion
disk viewed under an inclination of approximately 20\deg. The hydrogen
lines can be reproduced with a stellar-wind model which has a density
enhancement close to the disk's surface. This could point to the
presence of a disk wind.

Spectropolarimetric observations of the
H$\alpha$ emission show no difference in linear polarization with
respect to the continuum (i.e. line effect).
This suggests a spherical or circle-symmetric emission region
\citep{oudmaijer99}, which supports the low inclination found by
\citet{malbet07}. 


Recently, \citet{manoj07} have presented 1.3 millimeter (mm)
interferometric observations of MWC~297. They show that the mm
spectral slope of the spectral energy distribution (SED) is indicative
of grain growth in the circumstellar environment, or a very compact
structure which is optically thick at mm wavelengths. 
\citet{manoj07} also study the CO emission around MWC~297. They find an
extended optically thick CO cloud at $v_\mathrm{LSR} \approx 10$~\kms which
is resolved out in their interferometric observations. No compact CO
emission region, associated to the mm-continuum of the
MWC~297 system, is found. The authors interpret the latter as evidence
of strong depletion of CO in the disk.


\section{Observations}

\subsection {Interferometry}

MWC~297 was observed in the near- and mid-infrared with the
interferometric instruments AMBER \citep{petrov07} and MIDI
\citep{leinert03}. Both instruments are 
mounted on the VLTI\footnote{http://www.eso.org/projects/vlti/}. The
observations were performed with the 
1.8m Auxiliary Telescopes (ATs) in different baseline
configurations. In this paper, we use the VLTI nomenclature to refer
to the different telescope stations and baseline settings.

AMBER simultaneously combines the light of three telescopes and
currently 
operates in H- and K-band (1.6--2.5~$\mu$m). The fringes are spectrally
dispersed. Our data were taken in the low spectral resolution mode
(R$\sim$30, LR$-$HK mode). The data reduction was performed according
to the methods described in \citet{tatulli07}. Each observation
consists of a number of frames, i.e. exposures with short integration
times. We have tested different frame-selection algorithms 
(e.g. the suppression of frames with low photon counts in the
photometric channels; the selection of frames with statistically
significant piston estimates; the selection of frames with high
signal-to-noise visibility estimates), which were applied to both the
target and the calibrator star. Typically, 60--100\% of the frames
were selected. Little difference is seen between the resulting
calibrated visibilities, which confirms the excellent quality of the
AMBER data. Furthermore, we have repeated the AMBER observations on
the same baselines to check their consistency. The results of both
--independent--  
observations are equal within the error bars. We have averaged both
measurements to obtain the final set of AMBER visibility data.

Thanks to the combination of 3 telescopes, AMBER also provides a
spectrally dispersed closure phase (CP) measurement. To estimate the
uncertainty on the latter, we take the standard deviation of the closure
phase measurement of the calibrator, which is expected to have a zero
closure phase. The closure phase of MWC~297 is within the error bars
equal to zero ($|$CP$| < 5$\deg). Brightness distributions on the sky
that display 
deviations from centro-symmetry should always display a non-zero
closure phase, except at specific combinations of positions in the
(u,v)-plane where phases happen to sum up to zero. This is also true
if the baselines in the triangle are aligned, which is the case for
our observations. Hence, the zero 
AMBER closure phase strongly suggests the circumstellar geometry
of MWC~297 is point-symmetric along the position angle of the
baselines, at the spatial resolution of our observations.

We have extracted the H- and K-band AMBER spectrum of MWC~297 from the three
photometric channels. The flux in each channel is summed over all
frames, and calibrated using the total flux in the corresponding
channel of the calibrator star. A photospheric model of the calibrator
star is applied to convert to flux units. Finally, the absolute flux
level of the AMBER spectrum is determined by multiplicative scaling to
the mean value of the Infrared Space Observatory Short Wavelength
Spectrograph (ISO$-$SWS) spectrum
\citep{benedettini01, ackeiso} in the overlap region between both
(2.36--2.5~$\mu$m). The error on the spectral shape of the
spectrum is determined from the rms of the three photometric
channels. We note that the aperture of the ISO$-$SWS instrument at the
shortest wavelengths is 14\arcsec~$\times$~20\arcsec, much larger than
the seeing-limited resolution of a 1.8m telescope. However,
ground-based photometric measurements are in agreement with the ISO flux
levels. This indicates that the fraction of extended emission on a large
scale is negligible, and that the scaling of the AMBER spectrum to the
ISO spectrum is allowed. 

\begin{figure*}
\centering
\includegraphics[width=\textwidth]{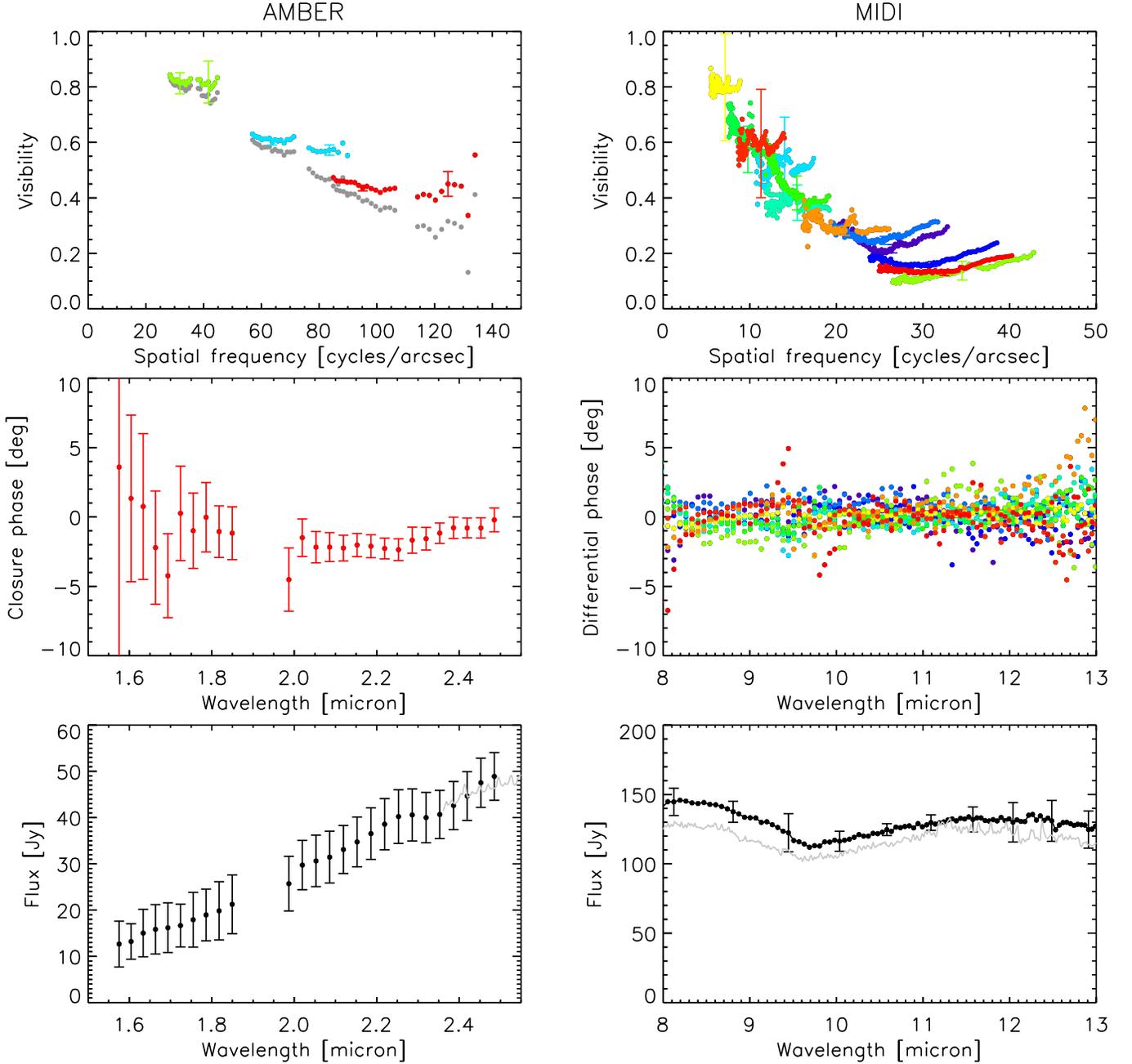}
\caption{Interferometric data. {\em Left, top to bottom:} the
  calibrated AMBER visibilities, closure phases and spectrum. {\em
  Right, top to bottom:} the calibrated MIDI visibilities, 
  differential phases and spectrum. Different colors refer to different
  baselines. 
  In the top left panel, the AMBER data
  corrected for the flux contribution of the central star are
  overplotted in grey (see Sect.~\ref{sif}). The grey line in the bottom
  panels is the ISO spectrum of MWC~297 \citep{ackeiso}. Representative
  error bars are overplotted, except for the MIDI differential
  phases; the average uncertainty on this quantity is 1 degree,
  consistent with the spread of the data around zero. [This figure is
  available in color in the electronic version.]} 
           \label{all_data.ps}%
 \end{figure*}

MIDI is a two-beam combiner which operates in N-band
(8--13~$\mu$m). The fringes are spectrally dispersed at a 
resolution of 30 (HIGH-SENS PRISM mode). We have used the
MIA$+$EWS\footnote{http://www.strw.leidenuniv.nl/$\sim$koehler/MIA+EWS-Manual/}
data reduction packages to obtain calibrated visibilties. The
so-called incoherent (based on power-spectrum analysis of the fringes)
and coherent 
(based on the information contained in the spectrally dispersed
fringes) data reduction results are 
consistent with each other, which indicates the good quality of the
data set. Furthermore, the coherent analysis 
provides differential phases. The latter are within the error equal
to zero on all MIDI baselines. This suggests that also the N-band
emission region, including the silicate feature, does not deviate
much from centro-symmetry.

The MIDI spectrum is determined in the following manner: The
raw spectrum of the science target is obtained by adding all
photometry files. It is flux-calibrated as described by
\citet{vanboekelphd} with the corresponding 
quantity of two calibrators, or two separate measurements of the same
calibrator, observed in the same night as MWC~297 and spaced in
airmass. Again, model photospheres were applied to compute the
theoretical spectrum of the calibrators. The angular extent of the
calibrators was used to scale the spectra to the right absolute flux
level. This procedure 
was obviously only possible in nights were more than one calibrator
was observed in the same MIDI setting. We obtain 9 independently
calibrated spectra. The final MIDI spectrum is the average of these,
with the error being the standard deviation. It is striking that the
independently calibrated ISO spectrum and MIDI
spectrum of MWC~297 have the same shape and flux level, consistent
with the IRAS 12~$\mu$m flux, despite the very
different aperture size or field-of-view of the respective
instruments. This argues in favor of a mid-IR emission region which is
spatially unresolved by a single telescope. The entire N-band
emission region is hence sampled by the MIDI interferometer.
This conclusion is confirmed by a 11.85~$\mu$m VISIR image of MWC~297,
in which the source is resolved, but on a much smaller scale
than observable with a 1.8m telescope (see Sect.~\ref{nmir}).

The log of the interferometric observations can be found in
Table~\ref{log}. The calibrator stars are indicated as well. The
resulting visibilities, closure phases and  
spectra are plotted in Fig.~\ref{all_data.ps}. The visibilities are
shown as a function of spatial frequency B/$\lambda$, with B the 
projected baseline length and $\lambda$ the wavelength of the
observations. The spatial frequency is a measure for the resolving
power obtained with a certain measurement.
To guide the reader who is not an interferometry expert:
each measurement at a certain baseline consists of a {\em spectrum}
of visibilities and phases, which not only sample a different
wavelength, but also a different spatial frequency {\em and} direction
on the sky. Visibility measurements at the same spatial frequency may
therefore differ due to a different baseline position angle and/or the
spectral dependence of the source's emission on the sky.
In the top right panel of Fig.~\ref{all_data.ps}, the MIDI
visibilities are presented. Here, it can be seen that the 10~micron
silicate feature leaves a spectral imprint on the visibilities. We
will come back to this in Sect.~\ref{rbd}.

\begin{table*}
\begin{minipage}[t]{\textwidth}
\caption{The log of the interferometric observations, sorted according
  to interferometric instrument and increasing baseline length. The
  first seven columns refer to the 
  observations of MWC~297, the last four to the calibrator's
  properties. As a reference, the wavelength-averaged calibrated
  visibility $<$V$>_\lambda$ is given. The flux of the
  calibrator is the 2MASS Ks-band magnitude 
  for AMBER observations and the IRAS 12~$\mu$m flux for MIDI. PB =
  projected baseline length; PA = projected baseline 
  angle.}
\label{log}
\centering
\renewcommand{\footnoterule}{}  
\begin{tabular}{cccccccccccc}
\hline\hline 
Night & UT & Instrument & Setting & PB & PA & Airmass & 
$<$V$>_\lambda$ & Calibrator & Sp.T. & Diameter & Flux \\ 
yyyy-mm-dd & hh:mm:ss & & & {\em m} & \deg~{\em E of N} &  &  &
& & {\em mas} & {\em mag} or {\em Jy} \\ 
\hline               
2007-04-14 & 08:09:13 & AMBER& E0--G0 & 14.2 & 68.4 & 1.16 & 0.78
&HD~166460& K2III & 1.40 & 2.58 \\
2007-06-19 & 07:10:31 & AMBER& E0--G0 & 14.9 & 73.0 & 1.20 & 0.86
&HD~166460& K2III & 1.40 & 2.58 \\
2007-04-14 & 08:09:13 & AMBER& G0--H0 & 28.5 & 68.4 & 1.16 & 0.60
&HD~166460& K2III & 1.40 & 2.58 \\
2007-06-19 & 07:10:31 & AMBER& G0--H0 & 29.9 & 73.0 & 1.20 & 0.59
&HD~166460& K2III & 1.40 & 2.58 \\
2007-04-14 & 08:09:13 & AMBER& E0--H0 & 42.7 & 68.4 & 1.16 & 0.46
&HD~166460& K2III & 1.40 & 2.58 \\
2007-06-19 & 07:10:31 & AMBER& E0--H0 & 44.4 & 73.0 & 1.20 & 0.41
&HD~166460& K2III & 1.40 & 2.58 \\
\hline
2007-04-12 & 08:22:50 & MIDI & E0--G0 & 14.3 & 68.5 & 1.15 & 0.79
&HD~148478 & M1.5I   & 25.92  & 3200 \\
2006-04-20 & 06:02:07 & MIDI & D0--G0 & 20.1 & 56.3 & 1.58 & 0.63
&HD~146051 & M0.5III & 9.83 & 150 \\
2007-04-17 & 06:50:30 & MIDI & G0--H0 & 23.4 & 61.9 & 1.37 &
0.60\footnote{The calibrators HD~82668 and HD~187642 were 
  observed at high airmass 
  ($\sim$2), which hampers a reliable calibration. We have averaged
  the calibrated visibilities. The large errors
  (Fig.~\ref{all_data.ps}) were deduced from the 
  standard deviation.}
&HD~82668  & K5III   & 6.99 & 73 \\
           &          &      &        &      &      &    &  
&HD~187642 & A7V     & 3.22 & 33 \\
2006-04-20 & 07:57:03 & MIDI & D0--G0 & 28.9 & 68.8 & 1.13 & 0.51
&HD~168454 & K3III   & 5.78 & 62 \\
2006-04-21 & 09:09:21 & MIDI & D0--G0 & 31.8 & 72.1 & 1.07 & 0.46
&HD~167618 & M3.5III & 11.33 & 214 \\
2006-04-22 & 09:59:19 & MIDI & D0--G0 & 31.9 & 73.1 & 1.09 & 0.38
&HD~168454 & K3III   & 5.78 & 62 \\
2007-04-19 & 08:00:32 & MIDI & E0--H0 & 43.3 & 68.8 & 1.14 & 0.30
&HD~187642 & A7V     & 3.22 & 33 \\
2006-05-23 & 05:06:38 & MIDI & A0--G0 & 52.6 & 66.0 & 1.23 & 0.27
&HD~167618 & M3.5III & 11.33 & 214 \\
2006-05-25 & 05:12:08 & MIDI & A0--G0 & 54.5 & 67.1 & 1.19 & 0.25
&HD~146051 & M0.5III & 9.83 & 150 \\
2006-05-25 & 07:20:53 & MIDI & A0--G0 & 64.0 & 72.7 & 1.07 & 0.18
&HD~146051 & M0.5III & 9.83 & 150 \\
2007-05-10 & 05:45:56 & MIDI & G1--H0 & 66.8 &172.9 & 1.27 & 0.14
&HD~167618 & M3.5III & 11.33 & 214 \\
2007-05-08 & 06:41:00 & MIDI & D0--G1 & 71.5 &129.5 & 1.14 & 0.13
&HD~167618& M3.5III & 11.33 & 214 \\
\hline               
\end{tabular}
\end{minipage}
\end{table*}

\subsection{Additional data}

\begin{figure}
\centering
\includegraphics[width=\columnwidth]{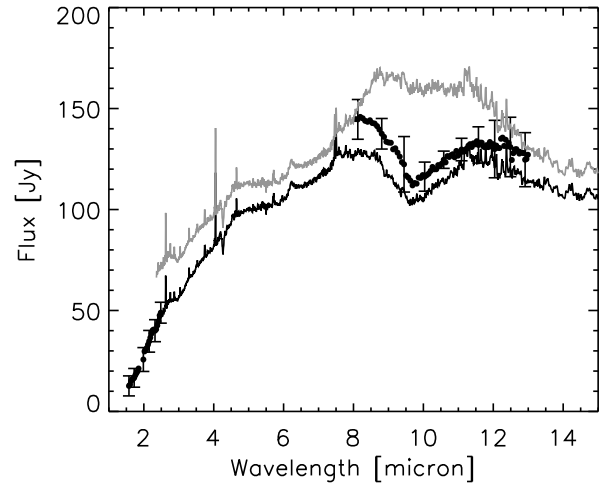}
\caption{ISO (black line), AMBER (dots) and MIDI (dots) spectra of
  MWC~297. The AMBER spectrum was scaled to the ISO
  spectrum in the overlap region between both spectra. The MIDI
  spectrum has 
  been calibrated independently and agrees well with the ISO
  spectrum. In grey, the ISO spectrum is shown, dereddened with the
  same extinction law as the optical photometry. A mild but significant
  silicate emission feature is present at 10 micron.}
           \label{all_spectra.ps}%
\end{figure}

We have expanded our data set with previously published photometric
and spectroscopic data. 
In Fig.~\ref{sed_w_BBs.ps}, the dereddened SED is shown, including the
ISO spectrum. The photometric data were taken from \citet{berrilli87},
\citet{mannings94a}, \citet{dewinter01}, and the 2MASS and IRAS
catalogues. The central star is modeled with a Kurucz 
model for a B1.5V star (T$_\mathrm{eff}$ = 25,400K, $\log g$ = 4.0).
The stellar radius, obtained from the fit of the photosphere
model to the UV and optical photometry, is 8$\pm$2~R$_\odot$. The R- and
I-band photometric points were excluded from this fit, since excess
flux is present at these wavelengths ($\sim$40\% and 20\%
respectively). The excess in R can be entirely attributed to the
extremely strong \Ha~emission in MWC~297. Also the I-band excess is
due to hydrogen emission lines. The Paschen lines measured by
\citet{andrillat98} account for the entire excess. We have applied
the interstellar dereddening law of 
\citet{savage79}, extended to infrared wavelengths with the law
described by \citet{steenman89, steenman91}. We set \Rv\ = 3.1 and
E(B-V) = 2.5, leading to \Av = 7.75, consistent with
\citet{drew97}. The AMBER, MIDI and ISO spectra are plotted in
Fig.~\ref{all_spectra.ps}. After dereddening, the ISO spectrum displays
a weak but significant 10~micron silicate emission feature. The
infrared excess luminosity L$_\mathrm{IR}$ is only 5\% of the
stellar luminosity L$_\star$.

In Section~\ref{ext}, we argue that the
extinction of the central star is most likely due to a foreground
interstellar molecular cloud, in between the observer and the MWC~297
system. Therefore, the {\em real} spectrum of the MWC~297 system is
the dereddened spectrum. In the following, we use the
dereddened AMBER, MIDI and ISO spectra, unless otherwise specified. We
stress that the dereddening has no direct consequence for the
interferometric measurements, since the latter are only
sensitive to the normalized brightness distribution of the source at
each wavelength.

Diffraction-limited images of MWC 297 in a broad-band filter centered 
at 11.9~$\mu$m, and intermediate-width filters centered on 11.3 and 
11.9~$\mu$m were obtained with the VLT Imager and Spectrometer for 
the mid-Infrared \citep[VISIR,][]{lagage04} during the night of 
April 13-14, 2007. Subtraction of the thermal emission from the sky,
as well as the telescope itself, was achieved by chop-nodding in the
North-South direction with a throw of 30\arcsec.
The effective integration time was 90 seconds in each
filter. Calibrator measurements in the same settings were obtained
shortly before the science observation. All observations were obtained
when the sources were close to zenit (airmass $\sim$ 1.05). The
full-width-at-half-maximum (FWHM) of the calibrators' broad-band image
is 0.33$\pm$0.01\arcsec, the FWHM of the MWC~297 image is 0.41\arcsec.

\begin{figure}
\centering
\includegraphics[width=\columnwidth]{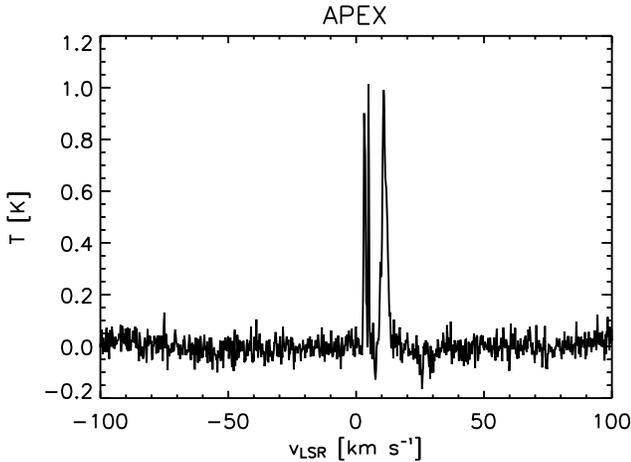}
\caption{APEX spectrum of the CO~$(3-2)$ emission line, averaged
over a 1\arcmin~$\times$~1\arcmin\ field around MWC~297. The three
detected features in the v$_\mathrm{LSR}$-range between 2 and 13~\kms 
are spatially extended and most likely produced in interstellar
clouds. No features exceeding the noise level are detected at any
other velocity.}
           \label{mwc297_apex.ps}%
\end{figure}

MWC~297 has been observed during a Science Verification run of the
Atacama Pathfinder Experiment (APEX\footnote{This publication is based
on data acquired with the Atacama Pathfinder Experiment (APEX). APEX
is a collaboration between the Max-Planck-Institut fur
Radioastronomie, the ESO, and the Onsala Space Observatory.}). The
APEX-2A heterodyne
receiver\footnote{http://www.apex-telescope.org/heterodyne/het345/}
mounted on this telescope provided a raster map of 
the CO~$(3-2)$ emission line at 
345.79 GHz in a 1\arcmin~$\times$~1\arcmin field around the
source, with a beam size of 17.5\arcsec. No spatial
dependence of the spectrum has been detected in the field, indicating
that the detected emission lines are due to a CO structure,
significantly larger than the mapped area. This is in agreement with
the results of \citet{manoj07}, who attribute this emission to
interstellar clouds. In Fig.~\ref{mwc297_apex.ps}, the
field-averaged spectrum is shown. No features are detected at high
red- or blueshift. The latter excludes the presence of a strong
CO outflow in MWC~297, on top of the non-detection of CO by
\citet{manoj07} in the circumstellar disk.

\section{Analysis}
%
%

\subsection{Simple geometric modeling \label{sif}}

We have fitted simple analytic models to the visibility data. Because
the AMBER closure phases and the MIDI differential phases are consistent with 
zero on all baseline settings, we have only applied point-symmetric
and concentric models. The brightness distributions in the \mbox{H-,}
K-, and N-bands 
may be dominated by different components. Therefore, we have 
first analyzed the data in the near- and mid-IR separately. In the
last part of this section, we confront the results and construct a
model which fits the interferometric observations and spectra in both
wavelength ranges simultaneously.

\subsubsection{Fit to the MIDI data}

Most of the MIDI data are obtained on baselines along the VLTI 0-line
(i.e. projected baseline angle $\sim$70\deg). 
We are hence predominantly sensitive to the N-band brightness
distribution projected onto this direction. The flux contribution of
the central star in the N-band is expected to be less than 0.5\% and
is therefore neglected in the fits described below.

For each observational wavelength, the MIDI visibility curve shows a
steady decline 
with increasing spatial frequency. Strong lobes are absent, although a
drop-off may be present at spatial frequencies around 30 cycles per arcsec. 
One-component uniform disk (UD), uniform ring (UR) and Gaussian disk
(GD) models cannot fit 
this shape. Two-concentric-component models (UD + point source, GD + point
source\footnote{Note that the point source in
these models does not represent the central star, but an unresolved
disk component.}, two UDs, two GDs) do better.  In general, the GD models do
better than the UD models (a factor two in reduced chi-square). In the
best-fit ring models, the inner gap is always smaller than 1 mas
(0.25~AU at 250~pc). The ring models are thus essentially equal
to the uniform-disk models. We have summarized the best-fit parameters
of some illustrative models in Table~\ref{midifits}. The amount of free
parameters in the two-component models was alternated. First, we have
fitted the extent of both components, and allowed for a
wavelength-dependent linear slope of the flux ratio (model type A in
Table~\ref{midifits}). The extent of the 
components was constrained to be the same over the 
N-band. Hence, this model is fully determined with only 4 parameters.
Second, the extent of the components is again
wavelength-independent, but now the flux ratio at each wavelength is a
free fitting parameter (model type B). The total number of parameters
is 2 plus the 
number of wavelength points. Finally, we have also fitted models
in which both the extent and the flux ratio are
wavelength-dependent (model type C). Models including point sources
have worse $\chi^2_\mathrm{red}$ values and are therefore excluded
from the table. The best model is a
2GD model with all parameters $\lambda$-dependent, with a
$\chi^2_\mathrm{red}$ close to 1. The best 2UD model {\em overfits}
the data ($\chi^2_\mathrm{red}$ = 0.6).

A number of authors \citep[e.g.][]{malbet07,manoj07} have suggested
that MWC~297 is surrounded by an accretion disk. Therefore, we have
fitted an alpha-disk (AD) model, in which it is assumed
that each annulus around the central star radiates as a blackbody. The
temperature at each distance from the star is parametrized with a
power law T(R)$\propto$R$^\alpha$. The six free parameters of this
model are the 
inner and outer radius, the temperature at the inner radius and the
power of the temperature law, the inclination of the disk and the
position angle of the major disk axis on the sky.
In a standard accretion disk, the temperature drops off with
radius proportional to R$^{-0.75}$. An accretion disk can therefore be
modeled with an alpha disk in which the power is set to $-0.75$.
The parameters of the best-fit AD models to the MIDI data are
summarized in Table~\ref{midiadfits}. The outer radius of the disk is
ill-defined by our data set, i.e. equally good fits can be obtained
with any value for this parameter. We have therefore fixed
R$_\mathrm{out}$ to the maximum outer radius determined by
\citet[][80~AU]{manoj07} from mm-observations. The AD models
produce a spectrum, which is constrained by the observed MIDI
spectrum. Reasonable
fits can be obtained, although the simple two-component
models presented in Table~\ref{midifits} appear to be much better.

\begin{table}
\begin{minipage}[t]{\columnwidth}
\caption{Results of some illustrative analytic model fits to the MIDI
  visibilities. The extent of the model components C1 and C2 is the
  wavelength-averaged FWHM for GDs and diameter for UDs. The
  wavelength-averaged flux ratio is included as well. \#p is the
  number of fitting parameters in the model, $\chi^2_\mathrm{red}$ the
  reduced chi-square value. \#$\lambda$ is the total number of
  wavelength bins in one MIDI observation. 
  See text for an explanation on the model types.
 }
\label{midifits} 
\renewcommand{\footnoterule}{}  
\centering   
\begin{tabular}{l c c c c c} 
\hline\hline     
Model & \multicolumn{3}{c}{Best-fit parameters} & \#p &
$\chi^2_\mathrm{red}$ \\
type  & Extent C1 & Extent C2 & Flux ratio & & \\
      & {\em mas} & {\em mas} &    C2/C1   & & \\
\hline      
2 UDs A & 65   & 23   & 0.74 &             4 & 4.2  \\  
2 GDs A & 38   & 10   & 0.28 &             4 & 2.8  \\  
2 UDs B & 65   & 23   & 0.79 & \#$\lambda$+2 & 3.6  \\  
2 GDs B & 40   & 12   & 0.38 & \#$\lambda$+2 & 1.9  \\  
2 UDs C & 96   & 32   & 1.44 & 3~$\times$~\#$\lambda$ & 0.6  \\  
2 GDs C & 79   & 20   & 1.35 & 3~$\times$~\#$\lambda$ & 1.1  \\  
\hline
\end{tabular}
\end{minipage}
\end{table}

\begin{table}
\begin{minipage}[t]{\columnwidth}
\caption{Results of alpha-disk (AD) fits to the MIDI
  visibilities. The alpha disk is determined by the inner and
  outer disk radius R$_\mathrm{in}$ and R$_\mathrm{out}$, the disk
  inclination $i$ and position angle PA, and the temperature law
  T~=~T$_0$\,(R/R$_\mathrm{in}$)$^{\alpha}$. \#p is the number of fitted
  parameters in the model, $\chi^2_\mathrm{red}$ the reduced
  chi-square value of the visibilities. 
  }
\label{midiadfits} 
\renewcommand{\footnoterule}{}  
\centering   
\begin{tabular}{l l l c c} 
\hline\hline     
Model & \multicolumn{2}{c}{Best-fit parameters} & \#p &
$\chi^2_\mathrm{red}$ \\
\hline      
AD      & T$_0$ = 1510 K             & $\alpha$ = $-$0.68       & 5  & 4.9 \\  
        & R$_\mathrm{in}$ = 4 mas    & $i$ = 40\deg &    &     \\  
        & R$_\mathrm{out}$ = 320 mas\footnote{Fixed a priori.} & PA = 160\deg &    &     \\  
\hline
Standard & T$_0$ = 1650 K             & $\alpha$ = $-$0.75$^a$ & 4  & 6.9 \\  
accretion& R$_\mathrm{in}$ = 4 mas    & $i$ = 45\deg &    &     \\  
disk     & R$_\mathrm{out}$ = 320 mas$^a$ & PA = 150\deg &    &     \\  
\hline
\end{tabular}
\end{minipage}
\end{table}

The main conclusion in this section is that simple analytic
two-component models do a much better job in reproducing the
observed MIDI visibilities than single-component models, a uniform
ring or an alpha disk. Two circumstellar structures are
present in the MWC~297 system; a smaller structure with a radius of 
$\sim$10~mas (2.5~AU at 250~pc) and a larger structure of $\sim$40~mas
(10~AU). Note that both components are extremely compact, given the
strong radiation of the central B1.5V star. In fact, both structures
reside entirely within the 1500K dust sublimation radius of 15~AU for
small (0.1~$\mu$m) amorphous olivine grains.

\subsubsection{Fit to the AMBER data}

Our observations in the H- and K-band are sensitive to the hottest
circumstellar matter present in the MWC~297 system. At these
wavelengths, also the central star is a contributor to the total
flux. The stellar radius of 8~R$_\odot$ corresponds to an angular
radius of 0.15~mas. At the
spatial resolution of our AMBER observations, the star is therefore
completely unresolved and can be treated as a point source.

Supported by the zero closure phase, we assume that the photocenter of
the circumstellar emission coincides with the central star. For each
wavelength bin, we can then 
extract the visibility $V_\mathrm{csm}$ of the circumstellar matter
from the measured visibility $V_\mathrm{meas}$ using 
\[ V_\mathrm{meas} = \frac{F_\star\,V_\star +
  F_\mathrm{csm}\,V_\mathrm{csm}}{F_\star + F_\mathrm{csm}},  \] 
with $V_\star = 1$, and $F_\star$ and $F_\mathrm{csm}$ the
(uncorrelated) flux of the star and the circumstellar matter
respectively. The latter are computed from the AMBER spectrum and the
photospheric model of the central star that fits the SED. In H-band,
the stellar flux 
contribution adds up to 20\%, in K to 10\%. Hereafter, we refer to
$V_\mathrm{csm}$ as the AMBER visibilities, unless otherwise
indicated.

Simple analytic models were fitted to the AMBER data. In
Table~\ref{ambfits} we summarize some illustrative results. All models
have wavelength-independent parameters, apart from the initial
correction of the visibilities
for the star-to-disk flux ratio. Increasing the complexity of the
models, e.g. by allowing a flux ratio of the components that
varies with wavelength, only decreases the quality of the fit in terms
of reduced chi-square due to the small number of AMBER data
points. We note that the best-fit UR models prefer a disk 
which has no inner gap (R$_\mathrm{in} < 0.2$~mas $\approx$
R$_\star$). As a result of this, the UR and UD fits are 
basically the same, with comparable $\chi^2_\mathrm{red}$
values.
It is remarkable that two-component models, with a fixed extent and
flux ratio over the H- and K-band, explain the AMBER data very well
within the observational errors.
A fit with an alpha-disk model to the AMBER data alone is not
possible, given the small number of AMBER points, and the fact that
all AMBER baselines are aligned: the inclination and position angle of
the disk are not constrained, the other parameters vary over a
broad range in parameter space. The parameter
$\alpha$, however, can be restricted based on the slope of the
visibility curve with wavelength. It appears to lie preferentially in
the range between $-$0.6 and $-$0.5, instead of around the standard
value for an accretion disk ($-$0.75).

\begin{table}
\begin{minipage}[t]{\columnwidth}
\caption{Results of some illustrative analytic model fits to the AMBER
  visibilities, corrected for the flux contribution of the central
  star. The nomenclature is the same as in
  Table~\ref{midifits}. See text for details.}
\label{ambfits} 
\renewcommand{\footnoterule}{}  
\centering   
\begin{tabular}{l c c c c c} 
\hline\hline     
Model & \multicolumn{3}{c}{Best-fit parameters} & \#p &
$\chi^2_\mathrm{red}$ \\
type  & Extent C1 & Extent C2 & Flux ratio & & \\
      & {\em mas} & {\em mas} &    C2/C1   & & \\
\hline      
UD+P    & 14.6 & --   & 0.77 & 2            & 2.5 \\
UR+P    & 0.2  & 7.3  & 0.44 & 3            & 2.5 \\
GD+P    & 8.3  & --   & 0.46 & 2            & 1.6 \\
2 UDs   & 23.2 & 7.3  & 3.6  & 3            & 0.9 \\
2 GDs   & 14   & 4.3  & 2.8  & 3            & 0.7 \\    
\hline
\end{tabular}
\end{minipage}
\end{table}

\subsection{Combining AMBER and MIDI}

\subsubsection{Disk models}

The major advantage of the combination of near-IR AMBER and mid-IR MIDI
data is the sensitivity to the temperature gradient of the
circumstellar matter. Analytic alpha-disk models can
therefore be subjected to a constraining test. The models are not only
constrained by the visibilities, but also by the spectra in H, K and N.
The best-fit standard accretion
disk ($\alpha$ fixed to $-0.75$) has a reduced chi-square of more than
60 and can therefore be ruled out. Leaving the power of the
temperature law free, the best fit is obtained with R$_\mathrm{in}$ =
0.42$\pm$0.06~AU = 1.7$\pm$0.2~mas, R$_\mathrm{out}$ = 20$\pm$20~AU =
80$\pm$80~mas, a temperature at R$_\mathrm{in}$ of 1600$\pm$250K and a
power $\alpha$=$-$0.55$\pm$0.03. The inclination and position angle of
the best-fit alpha disk are $i$=20$\pm$20\deg\ and PA=180$\pm$50\deg
respectively. The latter, as well as the outer radius, are
ill-defined by the data set. The reduced chi-square of this model is
8.8. The uncertainties on the best fit parameters are derived from
fits to virtual data sets, Monte-Carlo simulated around the original
data. 

Alpha-disk models fail to reproduce the
observations, because the presence of an inner gap is needed in the
models to keep the temperatures in the inner disk physical (i.e. not
exceeding a few 10,000K). The
visibilities show however that no gap in the 
brightness distribution is present at the resolution of our
observations. No secondary lobes are observed in the visibility
curves, which is e.g. witnessed by the tendency of the uniform ring
model fits to have an inner radius close to the stellar surface. Also
passive disk models with an inner disk gap
\citep[][and subsequent papers]{dullemond02}, in which
the central star is the only heating source of the circumstellar
matter, can a priori not explain the observations. If a hypothetical
passive disk around MWC~297 is seen under a moderate inclination
(0\deg\ $< i <$ 70\deg), the inner rim at dust sublimation temperature
would dominate the near- to mid-IR brightness distribution. Given the
large sublimation radius, a detectable gap would show up. If the
system is seen edge-on, its appearance on the sky changes. Although
present, a physical gap in the disk would not be detected in the
visibilities. 

\subsubsection{A dusty outflow or halo?}

%

It is suggested that the spectral characteristics of young stellar
objects can sometimes be explained without the need for a disk-like
geometry but instead assuming a spherical distribution of the dust
\citep[e.g.][]{vinkovic06}. Given the unclear origin of the
circumstellar material around MWC\,297, we deem it necessary to test
this hypothesis. To this end, we used the radiative transfer code
MoDust \citep{bouwman00,bouwmanphd} with optical constants from the
Jena
database\footnote{http:$//$www.astro.uni-jena.de$/$Laboratory$/$Database$/$databases.html}. 
\\ 

{\em A dusty outflow}

Mass loss through a dusty wind such as those found around evolved
stars is an a priori improbable source of the IR excess, as the
observed features are crystalline rather than amorphous, with a large
continuum contribution suggesting the presence of at least
micron-sized grains. Our attempts to model SED and visibilities
simultaneously with an outflow model failed in all possible
approaches, regardless of the dust species and grain characteristics
used: (1) An outflow model matching the SED
characteristics\footnote{Due to its large inner radius, this model
hardly obscures the photosphere, showing that a low circumstellar
A$_{\rm{V}}$ alone does not necessarly require a disk-like dust
distribution.} can be found, with an inner dust radius compatible with
its condensation temperature (R$_\mathrm{in} = 90\,R_{\star} =
3.5~\mathrm{AU}$, T$_\mathrm{in} = 1500$\,K), but the derived
visibilities are orders of magnitude too 
small at all wavelengths.  (2) Focussing instead on the observed
spatial dimensions of the circumstellar structure, it is possible to
reproduce the visibilities in the mid-IR in 2 ways: either with a
small inner dust radius or starting from the SED-matching model and
reducing the outflow to a thin shell at the inner radius. In the first
case, the dust which is unshielded from the stellar radiation reaches
unrealistic temperatures of 3000\,K and, moreover, this hot dust
generates such a strong near-IR excess that the AMBER visibilities can
not be reproduced. This is the case even when using only mm-sized or
oxide grains which resist heating much more than smaller and/or
silicate grains. In the second case, the SED is no longer matched, and
the dust which is at condensation temperature is still too bright
and/or resolved in the near-IR. (3) No way was found to reproduce the
AMBER visibilities, even without the constraint of the SED and MIDI
visibilities,  with anything remotely resembling a dusty outflow. \\

{\em Other density distributions}

We attempted fits with other density distributions as well, using
power laws ($\rho (R) = \rho_0 \,R^{\beta}$ with $\rho$ the dust
density, $R$ the distance from the central star and $\beta$ the free
parameter determining the density gradient) with $\beta$ ranging from
-3 to +3. Again, the inner and outer radius were varied but all to no
avail: requiring radiative equilibrium for the dust makes it
impossible to have it at a temperature of only 500\,K (the brightness
temperature of the larger component, and the temperature preferred for
that component by the analytical models presented in
Sect.~\ref{rbd}) so close to the central star without some
form of effective shielding. All fitted models have reduced chi-square
values over 100.

We conclude from these attempts that a dusty halo, be it an outflow or
some other density distribution, can explain none of the components
observed with AMBER and MIDI, in spite of good spectral agreement. 

\subsubsection{Approximating the brightness distribution \label{rbd}}

No tested physical model (accretion disk, halo, passive disk)
is able to explain the observations. Therefore, we aim at
approximating the near- to mid-IR brightness distribution of the
source using a combination of analytic models. This
approach focusses on the fitting of all interferometric data as a
whole and, as a by-product, delivers an {\em image} of the source
close to its real  appearance on the sky. One should however keep in
mind that the analytic components used in the fits do not necessarily
have physical meaning. They are merely building blocks for our
image. We will come back to this in Sect.~\ref{discussion}.

Two component models fit the AMBER and MIDI data well within the
errors. The AMBER visibilities, corrected for the contribution of the
unresolved central star of the system, are flux-dominated by a small
component. In the MIDI data, on the other hand, the most extended
component is dominant. The extent of the largest component in the
AMBER fits, and 
the smallest component in the MIDI fits, appears to be of the same
order. Furthermore, the closure and differential phases, in HK- and
N-band respectively, suggest a point-symmetric source geometry. 
Inspired by the AMBER and MIDI fits, we have built a concentric
three-component model, consisting of three Gaussian disks (3GD), to
fit all visibilities simultaneously. We assume that two of these
components are hot enough 
to radiate in the near-IR, and that the third one represents the
cooler material, which is responsible for the 10~micron silicate
feature and does not contribute to the AMBER flux. Since the
brightness temperatures of the two HK-band 
components is high (as expected from the AMBER fits), the N-band flux
contribution of the two hot components in the 3GD model is represented
by the Rayleigh-Jeans tail of a blackbody (i.e. F$_\nu \propto
\lambda^{-2}$). This approximation is made to suppress the number of 
model parameters. In summary, the model has 7 free 
parameters. The FWHMs of the GDs 
constitute three of those. Over the AMBER wavelength range 
(1.6--2.5~$\mu$m), we allow for a linear slope of the flux ratio of 
the two hottest components, implying two additional model parameters.
The last two parameters control the absolute flux levels of the
blackbody tails of the two hottest components at 10 micron. The flux
contribution of the coolest component is then fixed to the N-band
spectrum minus these blackbody contributions. At this stage, we do not
allow for asymmetry due to inclination. This will be discussed later
in this Section.

We fit this model to all visibility data simultaneously. The best-fit
parameters and errors are summarized in the upper part of
Table~\ref{bestfit3gd}. The three GD components were labeled A, B and
C from small (hot) to large (cold). The parameters are the FWHM of
each component, the linear flux ratio (FR) of component A and B in the
near-IR, and the absolute flux levels (FL) of components A and B at
10~$\mu$m. The
uncertainties on the best fit parameters are derived from fits to one
hundred Monte-Carlo simulated data sets, around the original data. 

\begin{table}
\begin{minipage}[t]{\columnwidth}
\caption{
  The parameters of the best-fit three-GD model. {\em Top:} Fit to all
  interferometric data simultaneously. {\em Middle:} Fit to subset
  only. {\em Bottom:} Inclination and position angle (PA) of the
  disk's
  major axis, based on the visibility measurements obtained on
  the two baselines with different baseline position angles. The
  reduced chi-square $\chi^2_\mathrm{red}$ for each fit is given. See
  text for details.
} 
\label{bestfit3gd} 
\renewcommand{\footnoterule}{}  
\centering   
\begin{tabular}{l l l l} 
\hline\hline     
\multicolumn{4}{c}{3GD fit to all} \\
\hline      
 FWHM &  A &       & 4.3$\pm$0.2 {\em mas} \\
 FWHM &  B &       &12.3$\pm$0.5 {\em mas} \\
 FWHM &  C &       &41.6$\pm$0.7 {\em mas} \\
 FR   & A/B& slope        & -1.0$\pm$0.4 {\em $\mu$m$^{-1}$} \\
      &    & at 2~$\mu$m  & 2.6$\pm$0.4  \\
 FL   &  A & at 10~$\mu$m &  2$\pm$2 {\em Jy} \\
 FL   &  B & at 10~$\mu$m & 45$\pm$2 {\em Jy} \\
 $\chi^2_\mathrm{red}$ & & & 4.5 \\
\hline      
\multicolumn{4}{c}{3GD fit to subset} \\
\hline      
 FWHM &  A &       & 4.2$\pm$0.3 {\em mas} \\
 FWHM &  B &       &  12$\pm$1 {\em mas} \\
 FWHM &  C &       &40.8$\pm$0.7 {\em mas} \\
 FR   & A/B& slope        & -0.9$\pm$0.4 {\em $\mu$m$^{-1}$} \\
      &    & at 2~$\mu$m  & 2.4$\pm$0.6  \\
 FL   &  A & at 10~$\mu$m &  8$\pm$5 {\em Jy} \\
 FL   &  B & at 10~$\mu$m & 36$\pm$5 {\em Jy} \\
 $\chi^2_\mathrm{red}$ & & & 4.1 \\
\hline
 $i$  &    &              & 40$\pm$10\deg \\
 PA   &    &              & 120$\pm$50\deg \\
 $\chi^2_\mathrm{red}$ & & & 3.8 \\
\hline
\end{tabular}
\end{minipage}
\end{table}

\begin{figure*}
\centering
\includegraphics[width=\textwidth]{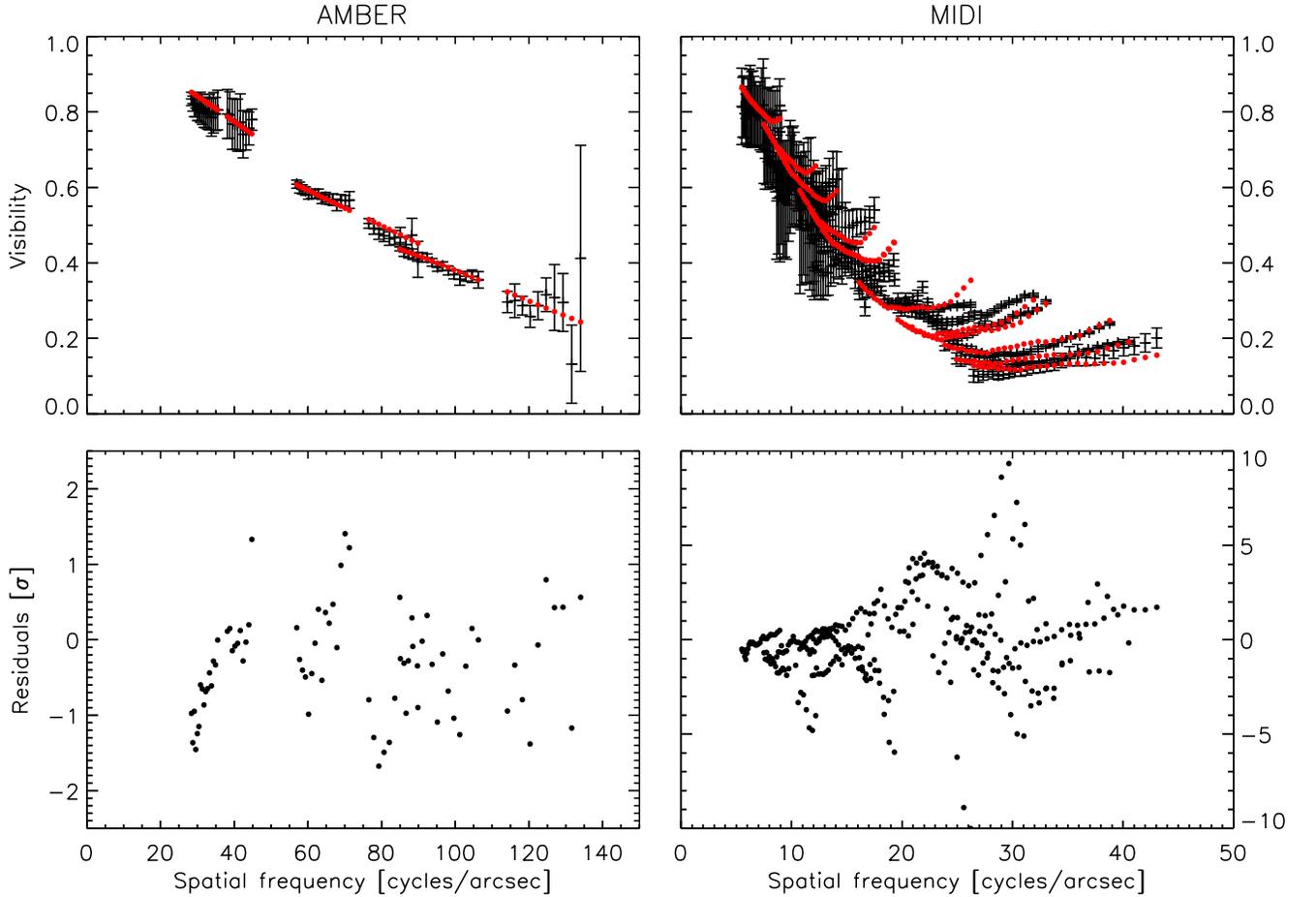}
\caption{{\em Top panels:} Best model fit to the visibility
  data. The AMBER data are corrected for the stellar flux
  contribution. The two longest MIDI baselines have different baseline
  position angles. The model fit to the visibilities related to the
  latter is the best-fit model which includes the effect of the disk's
  inclination, as described in the
  text. {\em Bottom panels:} Residual difference between
  observed and modeled visibilities, expressed in terms of the
  error $\sigma$ on the individual measurements. [This figure is
  available in color in the electronic version.]}
           \label{best_model.ps}%
 \end{figure*}

In our data set, two baselines have significantly different position
angles than the rest (PA=170\deg\ and 130\deg\ versus
PA$\approx$70\deg\ respectively). If present, strong asymmetries in
the brightness distribution should be observable at these
baselines. Therefore, we 
have repeated the fit on a smaller data set (hereafter called the
{\em subset} of data), limited to the data obtained at
PA$\approx$70\deg. Also these 
best-fit parameters are included in the middle part of
Table~\ref{bestfit3gd}. We note 
that the reduced chi-square is noticably better ($\chi^2_\mathrm{red} =
4.1$ versus 4.5, respectively), although only 85\% of the
data was used, and the number of free model parameters remained the
same. The extent of the three Gaussian disk components are comparable
in the two fits, and are 4, 12 and 41~mas respectively. This
corresponds to 1, 3 and 10~AU at a distance of 250~pc.

The blackbody brightness (BB) temperatures of the components are
deduced from their relative flux contributions to the AMBER+MIDI
spectrum. The errors are determined from the rather large errors on
the absolute flux level on the spectra, and the uncertainties on the
model parameters. The smallest component has a
BB temperature of 1700$\pm$200K, the middle component
930$\pm$80K. As a consequence of our modeling approach, 
the spectrum of the largest component contains the 10~micron silicate
emission feature, as well as part of the underlying continuum. 
We have determined the
temperature of this component, assuming that its flux is exclusively
produced by 0.1~$\mu$m olivine grains. The best fit was obtained with
a temperature of 520$\pm$20K and a column density larger than
0.02~g\,cm$^{-2}$. The latter value, in combination with the spatial
extent of the component, corresponds to a mass in small 
olivine grains of at least $2 \times 10^{-7}$~\Msun.

To get a handle on the possible elongation of the source's 
brightness distribution -- e.g. due to inclination -- we have taken
the 3GD model that fits the subset of data, and applied it to the
visibility data obtained on the two baselines with different position
angles. We insert one additional fitting parameter, a multiplicative
factor that can change the extent of the three components
simultaneously depending on baseline position angle. All relative
quantities (FWHM ratios, 
flux ratios) were fixed to the best-fit parameters, while the absolute
scaling of the extent of the model was fitted. This method yields
a factor at all position angles, which is nominally equal to 1 for
the subset of data 
at PA 70\deg, and 1.16$\pm$0.01 and 1.23$\pm$0.03 for the data along
PA 170\deg\ and 130\deg. Assuming that the circumstellar structure
around MWC~297 is a geometrically flat disk -- supported by
the low L$_\mathrm{IR}$ / L$_\star$ ratio of 5\% -- these numbers can
be translated into an inclination and position angle of the major disk
axis on the sky of 40$\pm$10\deg\ and 120$\pm$50\deg\
respectively. The extent of the components along the disk major axis
is 1.3$\pm$0.1 times larger than along PA=70\deg. The reduced
chi-square of the {\em rescaled} model visibilities, compensated for the
additional model parameter, is 3.8 (bottom part
Table~\ref{bestfit3gd}). Although a large inclination is 
excludable, we caution against overinterpretation of the inclination
and disk position angle estimates. The two baselines with different 
PAs are also the longest baselines, and hence sample a slightly
different spatial-frequency domain than the subset. A
perfectly 
spherically symmetric model may therefore fit the data just as
well (as was e.g. shown with the MIDI fits). To straighten out the
issue and determine the exact inclination, more visibility
measurements at comparable spatial frequencies, but different
baseline angles, are required. 

\begin{figure}
\centering
\includegraphics[width=\columnwidth]{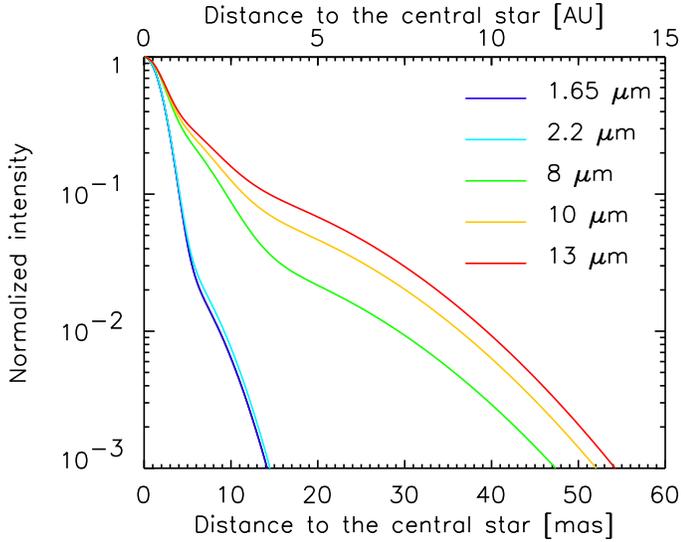}
\caption{Radial cut through the normalized intensity distribution of
  the best-fit model. Different curves refer to cuts at different
  wavelengths: 1.65, 2.2, 8, 10 and 13~$\mu$m. The extent of the model
  increases with increasing wavelengths. The flux contribution of the
  central star is excluded from the plot for clarity. [This figure is
  available in color in the electronic version.]}
           \label{best_model_image.ps}%
 \end{figure}

We consider our fit to the subset of data as our most
reliable fit, since no issues related to the position angle are
present. The model fit to the visibility data is shown in
Fig.~\ref{best_model.ps}. In Fig.~\ref{best_model_image.ps}, a radial
cut of the intensity distribution of the model at different
wavelengths is shown. Note that, although the FWHM of the largest GD
component is 40~mas (10~AU), the FWHM of the {\em total} image is less
than 6~mas (1.5~AU) at all wavelengths. 
The extent of our approximated image of MWC~297 
is in perfect agreement with the VISIR image at 11.85~$\mu$m.
We have convolved the model
image with a Gaussian curve of FWHM 0.33$\pm$0.01\arcsec, to mimic the
point spread function estimated from the two calibrator
measurements. The resulting image has a FWHM of 0.39$\pm$0.01\arcsec,
in perfect agreement with the measured 0.41\arcsec.

\begin{figure}
\centering
\includegraphics[width=\columnwidth]{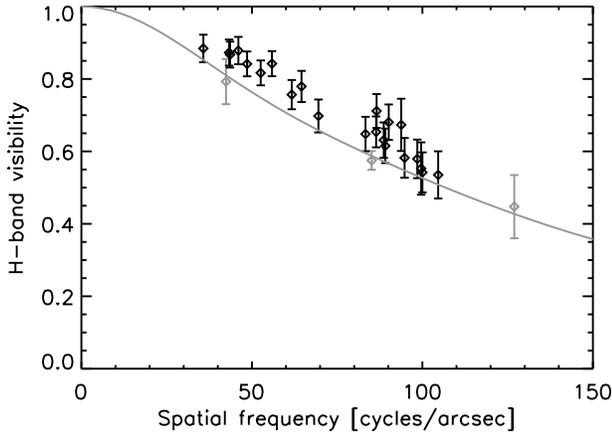}
\caption{IOTA broad-H-band data published by
  \citet{monnier06}. For comparison, the (narrow-band) AMBER
  visibilities at 1.65~$\mu$m are overplotted in grey, as well as our
  best-fit model converted to broad-band visibilities (solid line).}
           \label{iota_vs_amb.ps}%
 \end{figure}

We have compared the H-band AMBER and model visibilities to the
broad-band visibilities obtained by \citet{monnier06} with IOTA. In
Fig.~\ref{iota_vs_amb.ps}, their H-band visibilities are shown
together with the AMBER visibilities at 1.65~$\mu$m and the broad-band
visibilities of our best-fit model. The IOTA measurements are in
relatively good agreement with our data given the large error bars,
although they appear to be systematically higher. It is not clear to
us why this discrepancy is present. 
The IOTA H-band data were obtained at different baseline
position angles. The smooth decline of the visibilities with spatial
frequency, independent of position angle, shows that
the circumstellar emission region does not display strong
asymmetry. In case of a flat disk, this corresponds to a moderate
inclination of {\em at most} 40\deg, consistent with our upper limit. 

\subsection{Consistency with the SED}

\begin{figure*}
\centering
\includegraphics[width=\textwidth]{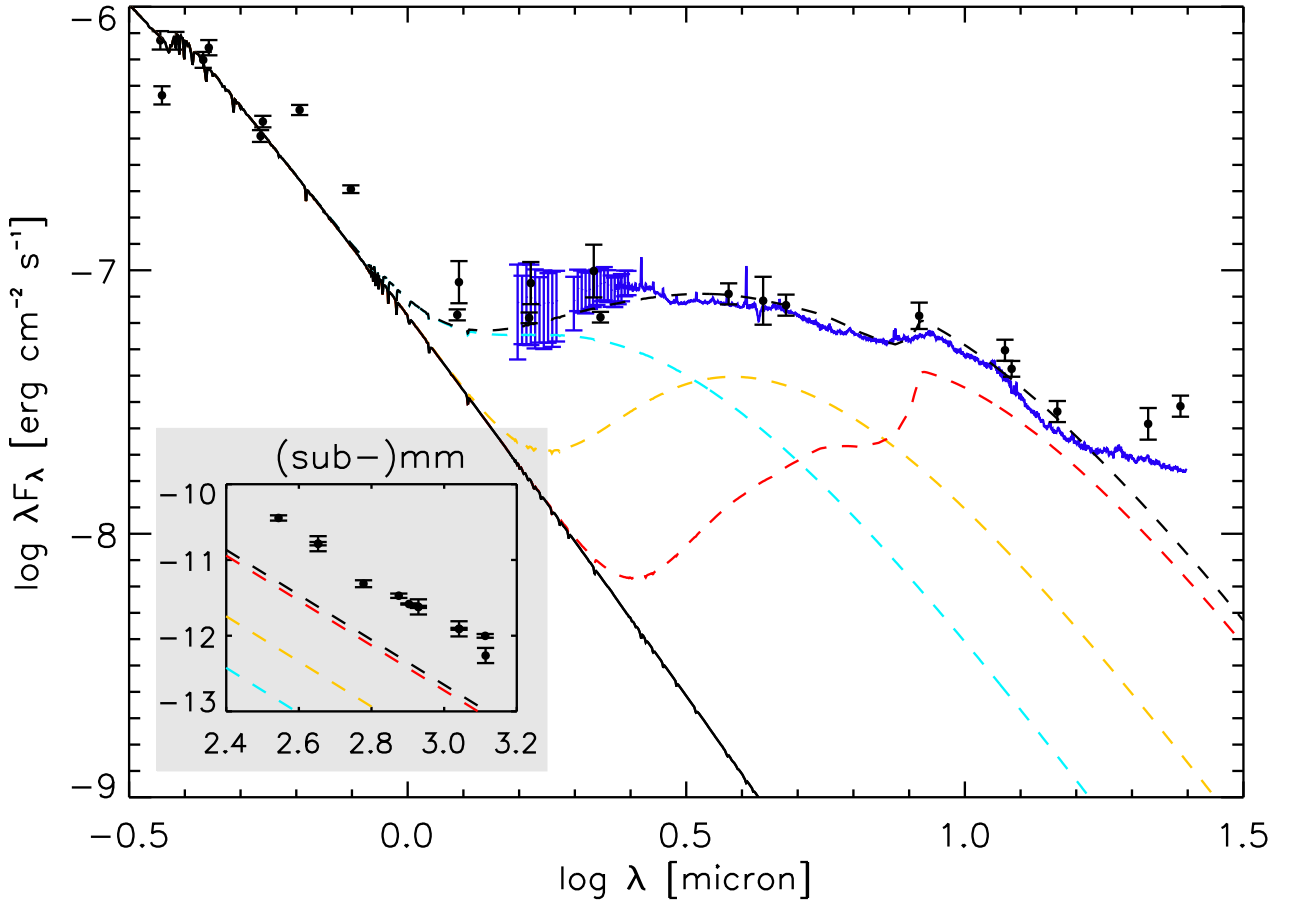}
\caption{Dereddened SED of MWC~297. Black dots with error bars are
photometric measurements. The full black
line is a Kurucz model of a 
B1.5V star (T$_\mathrm{eff}$=25400K, $\log g$=4.0), the full blue line
is the dereddened ISO spectrum. The blue error bars represent the
dereddened AMBER spectrum. The flux contribution of the three
Gaussian-disk components in the model are shown as dashed lines,
adding up to the black dashed line. {\em Inset:} At sub-mm/mm
wavelengths, the model underestimates the observed fluxes, indicating
the presence of an additional, cold component in the circumstellar
environment of MWC~297. [This figure is available in color in the
electronic version.]}
           \label{sed_w_BBs.ps}%
 \end{figure*}

The brightness temperatures derived for the three components in our
best-fit model, are consistent with the near- to mid-IR SED. The model
emission accounts for the total near- to mid-IR flux. In
Fig.~\ref{sed_w_BBs.ps}, the contribution of the components to the
total flux is shown. The spectrum of the largest component is a
single-temperature Gaussian disk consisting of olivines with a column
density of 0.02~g\,cm$^{-2}$. At sub-mm/mm wavelengths, only the
coolest component contributes significantly. The photometric points at
these wavelengths are clearly higher than the predicted model
flux. This points to the presence of an additional cold dust
component, which does not contribute flux at near- to mid-IR
wavelengths. Following \citet{manoj07}, 200~mJy of the 1.3~mm flux can
be attributed to dust emission, while the rest is free-free emission
from an ionized wind. After subtraction of the contribution of the
three warm near-/mid-IR components, 155~mJy remains. The combination
of the 1.3~mm flux and the fact that the cold dust does not contribute
to the mid-IR flux, places an upper limit of 200K on the brightness
temperature of this material, and a lower limit of 160~mas to its
extent. The limits set by spatially resolved mm observations
\citep[$<$~230~mas,][]{mannings94a, manoj07}, constrain the brightness
temperature to more than 100K. The brightness temperature of the cold
dust is therefore restricted to the interval between 100 and 200K.

\begin{figure*}
\centering
\includegraphics[width=\textwidth]{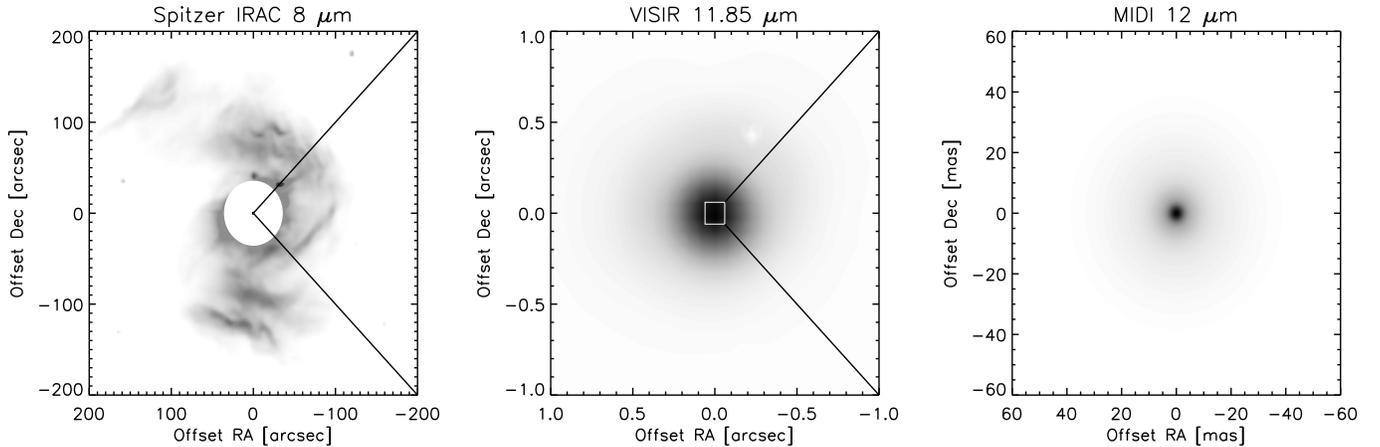}
\caption{Traveling to the center of MWC~297. {\em Left:} Spitzer
  IRAC image at 8~$\mu$m. Clearly visible is a complex, large-scale
  cloud structure. A numerical circular mask covers the saturated part
  of the image. {\em Middle:} Zoom-in at the center of
  the Spitzer 
  image: VISIR image at 11.85~$\mu$m. The image is seeing-limited, but
  its FWHM is significantly larger than that of the calibrators. {\em
  Right:} Zoom-in at the center of the VISIR image: our best-fit
  interferometric image at 12~$\mu$m. The greyscale in
  the three images is linear.}
           \label{MWC297_Nband_allscales.ps}%
 \end{figure*}

Spitzer IRAC images of MWC~297 show very complex cloud
structure on arcminute scales
(Fig.~\ref{MWC297_Nband_allscales.ps}). Its extent and structure
coincides with that of the 50-100~$\mu$m brightness 
distribution measured by \citet{difrancesco94}. This
cold material is likely responsible for the additional far-IR
($>$25~$\mu$m) flux excess, evident in e.g. the IRAS 25, 60 and
100~micron points. Also the ISO$-$SWS and ISO$-$LWS (Long
Wavelength Spectrograph) spectra, which are obtained with increasing
aperture size towards longer wavelengths \citep{benedettini01}, suffer
from this large-scale emission. It is possible that the mm
observations are partially affected by this emission as well, although
most of it would be resolved out. At near- and mid-IR wavelengths, the
flux contribution of the clouds is negligible: at 8 micron, the total
flux adds up to 130~Jy, spread out over roughly 10 square
arcminutes. Assuming a homogeneous emission region, the flux inside
the ISO 14\arcsec~$\times$~20\arcsec\ aperture would be of the order
of 1~Jy ($<$1\%).

\section{Discussion \label{discussion}}

\subsection{The extinction towards the MWC~297 system \label{ext}}

The flux of the central star in the MWC~297 system is
attenuated with 8 magnitudes in V. \citet{drew97} suggest that the
real extinction could be even stronger, if the observed stellar flux
is scattered, rather than direct light. However, this seems
unlikely; The application of an interstellar extinction law
\citep[][with \Av = 7.75 and \Rv = 3.1]{savage79, steenman89,
  steenman91}, makes that the 
dereddened photometry agrees with the expected fluxes of a B1.5V star at
250~pc from UV to the optical. In the case of scattering, the observed
photometry would have been bluer. Application of an interstellar
extinction law would in this case be inappropriate and would
yield results inconsistent with a B1.5V model.

A priori, it is unclear whether the attenuation of the central star is
due to circumstellar matter, physically linked to the central star, or
interstellar dust. However, in the case of MWC~297, the extinction
is most likely interstellar. An extended CO cloud is detected in the
direction of the 
target \citep{manoj07}. The extent of this cloud is at least
1\arcmin~$\times$~1\arcmin, much larger than the extent of the
near- and mid-IR emission region of MWC~297. 
The ISO spectrum of MWC~297 shows H$_2$O
and CO$_2$ ice absorption bands, indicative of a cold
environment. It is improbable that these ices are abundant in a small
(R~$<$~80~AU) and hence warm environment close to a B1.5V
star. Furthermore, attenuation of the central star can be descibed
with an interstellar extinction law, where small (0.1~$\mu$m) grains
are the main source of opacity.
The presence of small dust grains, CO$_2$ and H$_2$O ice, and the
large-scale CO emission are naturally explained by an interstellar
molecular cloud in between the observer and the MWC~297 system.
Moreover, the proximity (0.1~pc at 250~pc) of the large-scale cloud
structure present in the Spitzer images --which is also evident from
the extended 50-100~$\mu$m emission found by \citet{difrancesco94}--
strongly suggests that the latter is 
remnant material from the natal cloud from which MWC~297 and the
nearby lower-mass young stars \citep{vink05b,damiani06} were
formed. It is probable that the interstellar cloud in the
line-of-sight is part of this large structure surrounding the MWC~297
stellar group. 

The main consequence of this conclusion for the present paper, is the
fact that the entire MWC~297 system (i.e. the central star + the
circumstellar matter responsible for the near- to mid-IR flux) is {\em
screened} by this cloud. This is not important for the
interpretation of the interferometric data, because the latter are
only sensitive to the {\em normalized} brightness distribution on the
sky. Relative quantities (e.g. flux ratios between two components in
the interferometric models) can however be transformed into
absolute quantities (e.g. brightness temperatures), only if the {\em
unobscured} spectrum of the system is known. We have dereddened the
AMBER, MIDI and ISO spectra of MWC~297 for that purpose, with the same
extinction law extending from UV to mid-IR.

\subsection{The near- and mid-IR geometry of MWC~297 \label{nmir}}

Our investigation of the near- and mid-IR brightness structure of the
MWC~297 system yields the following results:

\begin{itemize}
\item MWC~297 is an extremely compact source. The bulk of the
  near- to mid-IR emission emanates from a region remarkably close to
  the central B1.5V star. The extent of the system in the \mbox{H-,}
  K- and N-bands (image FWHM $\sim 1.5$~AU) is 
  well within the maximal outer radius of the system determined
  from mm observations \citep[$<$~60$-$80~AU,][]{mannings94a,manoj07}.
\item The closure and differential phases obtained with 
  respectively AMBER and MIDI are equal to zero within the
  errors. There is no evidence of strong deviations from 
  centro-symmetry along a position angle of 70\deg\ E of N at the
  spatial resolution of our observations, neither in the continuum nor
  the silicate feature.
\item There is no significant IR-emission-free
  gap between the circumstellar matter and the stellar surface at the
  resolution of our observations.
\item Tentative evidence is found for a moderate elongation of the
  brightness distribution, or a moderate inclination
  of the circumstellar disk ($i < 40$\deg). A high
  inclination is excludable.
\item Parametrized accretion disk models cannot explain the data. This
  either suggests that the circumstellar matter around MWC~297 does
  not reside in an accretion disk, or that simple models are not
  appropriate to explain the characteristics at the small angular
  scales observed.
\item A spherically symmetric dusty outflow or halo cannot account for
  the interferometric and photometric observations either.
\end{itemize}

We have used geometrical models to approximate the circumstellar
brightness distribution of
MWC~297, with as few free parameters and analytic components as
necessary. Three circumstellar components suffice. However, it is
unclear what the physical meaning of these components is. Although it
might be that they are independent physical structures, it is more
likely that the three Gaussian disks sample the temperature gradient
of a smooth density distribution. Our analysis has
shown that a power-law approximation of the temperature is
inappropriate for the circumstellar matter around
MWC~297. Furthermore, the strong radial drop-off of the temperature
seems to suggest that the circumstellar matter is
optically thick in the radial direction. The halo modeling has shown
that 
an optically thin or moderately thick circumstellar medium produces a
much shallower temperature gradient: at the location of the largest
geometrical component, the temperature would be much higher than
observed. The most likely geometry of the circumstellar matter is
therefore a flattened disk, which is viewed under a
moderate inclination. However, this disk is significantly different
from a typical lower-mass Herbig~Ae or T~Tauri accretion disk: it is
compact and geometrically rather flat (as supported by the low IR
excess luminosity).

\subsection{On the nature of the circumstellar matter
  \label{nature}} 

The circumstellar structure in the MWC~297 system is remarkable in many
respects. We highlight some literature results and discuss these in
the framework of the present paper.

\begin{itemize}
\item The central star rotates with a projected velocity $\mathrm{v}
  \sin i = 350 \pm 50$~\kms \citep{drew97}. Critical rotation occurs
  when v$_\mathrm{rot}$ attains $\sqrt{2
    GM_\star/(3 R_\star)} = 450 \pm 70$~\kms, or equivalently when $i
  < 55 \pm 15$\deg. The 
  interferometric measurements suggest a moderate inclination $i < 40
  \pm 10$\deg, and hence that the star is indeed rotating at or close
  to critical velocity.
\item Extremely strong hydrogen emission lines from optical to mid-IR
  wavelengths are observed. The spectral line profiles are rather
  narrow (H$\alpha$ FWHM $\sim$ 200~\kms) compared to e.g. the
  half-maximum widths observed 
  in other Herbig Be stars. The extinction of the hydrogen lines is
  consistent with an \Av of 8 \citep{drew97}, which suggests that the
  hydrogen emission region is attenuated by the same material as the
  central star.
\item The forbidden \OI emission lines at 6300 and 6363\A are
  extremely strong, but also extremely narrow
  \citep[FWHM $\sim$ 22~\kms,][]{ackeoi}. The 
  lines are thermally excited, as proven by the detection of the
  \OI 5577\A line \citep{andrillat98}; The line ratio of the 6300 and
  5577\A lines is 5$\pm$1, combining the equivalent width measurements
  of \citet{andrillat98} and \citet{ackeoi}. Typical Herbig Be sources
  \citep[e.g. \object{Z~CMa},\object{CD$-$42$^\circ$11721},
  \object{PV~Cep}, \object{V645~Cyg};][]{ackeoi} display strong \OI
  lines with extended blue wings, up to  
  several hundreds of \kms. The latter is a strong indication for the
  presence of a bipolar outflow. If the \OI emission in MWC~297 also
  emanates from an outflow, the outflow axis should lie in the plane
  of the sky \citep{ackeoi}, to match the limited velocity dispersion
  of the observed spectral lines. Given the small disk inclination,
  however, it is most probable that the \OI emission emanates from the
  disk itself.
\item \citet{manoj07} find a shallow sub-mm slope of the SED,
  indicative for the presence of large grains, or an emission region
  which is optically thick at these wavelengths. The first would
  indicate grain growth, the former is consistent with a very compact 
  and dense circumstellar disk.
\item In the 3$-$micron range, the spectra of MWC~297 display diamond
  emission \citep{terada01,ackenano}. Diamonds form under conditions of
  severe temperature and pressure, both of which appear to be present
  in the circumstellar environment of MWC~297.
\item H$_2$O maser emission has been found by \citet{han98} at a
  v$_\mathrm{LSR}$ of $-75.6$~\kms. The emission is blueshifted by
  $\sim 85$~\kms~compared to the radial velocity of the central star
  (v$_\mathrm{LSR} = 10 \pm 2$~\kms, based on the peak position of
  optical emission lines). \citet{palla93} suggest that in Herbig Be
  stars the origin of this emission lies in a collimated (bipolar)
  outflow. The detection of water emission seems to suggest that also
  MWC~297 drives a molecular outflow, despite the absence of a
  detected CO outflow, even at high blueshifts
  (see Fig.~\ref{mwc297_apex.ps}). Given the strong 
  blueshift, the outflow axis of 
  the H$_2$O must be approximately aligned with the line of sight. The
  receeding (redshifted) emission would then be obscured by the
  circumstellar disk. This is consistent with a small disk
  inclination. 
\end{itemize}

MWC~297 is a high-mass young star, accompanied by lower-mass newborn
stars and remnants of its natal cloud. The central B1.5V star is a
critical rotator. It is surrounded by a dense gaseous and dust-rich
disk, which reaches in close to the stellar surface and produces the
prominent gaseous emission lines in the optical-to-IR spectrum. The
star+disk are viewed under a moderate inclination angle.

\section{MWC~297: a critical rotator and Be star progenitor
  \label{Be}}

Stellar rotation has, in recent years, been realised to be a
fundamental property that can determine the evolution and ultimate
fate of high-mass stars \citep{meynet00,maeder00}.  Rapid
rotation impacts on stellar life times and alters the chemical
evolution of the star. Knowing stellar rotation as a function of
mass and age is thus essential to understanding high-mass star
evolution. In our galaxy, some twenty per cent of all high-mass
stars, the classical Be stars, show rapid rotation and associated
mass loss. Our identification of MWC~297 as a critical rotator proves
that high-mass stars can achieve such high rotation rates at
birth. In this section, we discuss the consequences of this
discovery. Furthermore, we argue that MWC~297 is an evolutionary
progenitor of a classical Be star.

Observations have provided increasing evidence that
also high-mass stars form due to mass gain via an accretion disk
\citep{vink02}, as do solar-mass stars. For the latter,
the rotational velocity at birth is thought to be determined by the
interaction of the star and its accretion disk \citep{shu94}. Magnetic
fields lock the stellar 
surface rotation rate on the orbital rate close
to the inner edge of the disk. A high accretion rate, as well as a low
magnetic field strength, shifts the distance at which stellar and disk
rotation are locked to smaller radius, implicating a higher stellar
rotation rate. Therefore, theory predicts that stars with a small inner
disk radius have a high surface rotation rate at birth. 

The evolution of the equatorial rotation speed of high-mass stars is
governed by the combined effects of surface mass loss and the internal
redistribution of angular momentum. The first removes angular
momentum and is believed to be important for very massive
(M$>$25~\Msun) stars; its influence for lower-mass stars is unclear
\citep{ekstrom08} but can be significant. 
Angular momentum redistribution inside the star, however, can result
in an increase of the ratio between equatorial rotation speed and the
critical velocity. Theory 
predicts that the critical velocity may be attained for 
stars of 10$-$20~\Msun at a considerable age of 10 million
years. On the other hand, very little is known about the rotational
velocities of high-mass stars at birth, and theory is currently unable
to describe the evolution of stars born at the critical speed. It is
in this framework that the critical rotation of MWC~297 attains its
full significance.

Two observations imply that a dense and opaque
gaseous innermost disk, very close to the stellar surface, shields the
outer disk in the MWC~297 system. (1) The occurrence of dust so close
to the star is 
remarkable, as direct stellar irradiation would heat it there far
above its sublimation temperature. In lower-mass young stars, stellar
irradiation makes the innermost disk dust-free and opens up a gap
between star and dust disk \citep{monnier05}. No such gap is observed
in the disk 
of MWC~297. (2) Given the strength of the optical emission lines but
their small radial-velocity dispersion, these lines must mainly originate
in the disk, which is at low inclination. A large amount of
neutral oxygen and hydrogen 
atoms are thus present close to the star, which confirms that the
ionising stellar photons are absorbed efficiently close to the
star.

The fact that this remarkably compact disk encompasses a critically
rotating star suggests a causal link.  This link could be the
formation scenario discussed above. In that case, MWC~297 would either
have a weak magnetic field, have experienced high-rate mass accretion,
or a combination of both. It would mean that a scenario initially
developed for low-mass stars, also applies for their high-mass
counterparts. Accretion may still be ongoing in this case. The current
observations, however, do not allow for a definite judgement on
whether the matter in the innermost disk is moving inward or
outward. It cannot be excluded that accretion has ceased, and that the
critical rotation of the star currently induces mass loss to the disk
through an equatorial outflow. Both hypotheses are consistent with the
absence of an inner gap implied by the near- and mid-infrared
observations, and the 
presence of high-density gas very close to the stellar surface.

The discovery of a young star rotating at critical velocity
proves that stars can achieve such high rotation rates at birth.
This conclusion naturally raises intriguing questions on the initial
distribution of rotational rates in newly formed
high-mass stars. At this point, it is difficult to estimate the
frequency at which critically spinning stars are produced. 
A hint that this fraction cannot be too small may be found
in the proximity of MWC~297: it is the closest high-mass young star
known to date, and as such the prime target for investigation. That
this particular star rotates at critical velocity 
suggests that we were either very fortunate, or that fast rotation at
birth is more common than generally assumed. 

It is tempting to speculate about the future evolution of MWC~297.
The star will most likely evolve into a classical Be star. The latter
are high-mass main-sequence stars which exhibit line emission
produced in a gaseous equatorial disk: the line profiles of the
Balmer lines 
indicate Keplerian rotation and rather small outflow velocities.
Since classical Be stars share the common characteristic of
faster-than-average stellar rotation, it is likely that strong
equatorial centrifugal forces contribute to the origin of the disks.
However, observed rotation velocities do not reach the critical
velocity, hence the need to invoke additional effects, such as
pulsations or magnetic streams, to explain disk formation
\citep{porter03}. In a first phase, MWC~297's disk will evolve from
dust-rich to purely 
gaseous: The high density in the innermost disk will decrease, either
due to a decrease of the accretion rate, or a decrease of the
stellar equatorial mass loss.
The disk will eventually become transparant to stellar
photons, which will sublimate the dust grains. The neutral oxygen and
hydrogen will become 
ionised, implying the disappearance of the forbidden lines and
significantly reducing the strength of hydrogen emission. MWC~297 will
then have all the features of a classical Be star. 

Our study demonstrates the power of interferometry to determine the
initial rotation rates of massive stars.  When such studies can be
extended to a statistically significant sample, it will become
possible to assess whether MCW~297 is an exceptional object or whether
critical rotation is the general rule for newborn massive stars.  In
the latter case, the much slower rotation of most main-sequence stars
would imply that the processes which carry away angular momentum are
very efficient in our galaxy.  In the former case, MWC~297 would
qualify as a candidate for the hypothetical objects in which extreme
rotational mixing induces an almost homogeneous internal composition
throughout the main-sequence phase, hence causing the star to bypass
the red-supergiant phase \citep{maeder87}. MCW~297 would then be a
possible progenitor of a gamma-ray burster \citep{woosley06} close to
our sun.  However, gamma-ray bursters appear to occur 
preferentially in low-metallicity environments \citep{wolf07}, where
indeed the mass loss  and associated angular-momentum losses are
expected to be lower than in our galaxy \citep{macfadyen99}.

\acknowledgements{The authors want to thank J.~Monnier for kindly
  and quickly providing us with the IOTA H-band visibility data. 
  This publication makes use of
  data products from the Two Micron All Sky Survey, which is a joint
  project of the University of Massachusetts and the Infrared
  Processing and Analysis Center/California Institute of Technology,
  funded by the National Aeronautics and Space Administration and the
  National Science Foundation.}


\bibliographystyle{aa}
\bibliography{references.bib}

\begin{thebibliography}{47}
\expandafter\ifx\csname natexlab\endcsname\relax\def\natexlab#1{#1}\fi

\bibitem[{{Acke} \& {van den Ancker}(2004)}]{ackeiso}
{Acke}, B. \& {van den Ancker}, M.~E. 2004, \aap, 426, 151

\bibitem[{{Acke} \& {van den Ancker}(2006)}]{ackenano}
{Acke}, B. \& {van den Ancker}, M.~E. 2006, \aap, 457, 171

\bibitem[{{Acke} {et~al.}(2005){Acke}, {van den Ancker}, \&
  {Dullemond}}]{ackeoi}
{Acke}, B., {van den Ancker}, M.~E., \& {Dullemond}, C.~P. 2005, \aap, 436, 209

\bibitem[{{Andrillat} \& {Jaschek}(1998)}]{andrillat98}
{Andrillat}, Y. \& {Jaschek}, C. 1998, \aaps, 131, 479

\bibitem[{{Benedettini} {et~al.}(2001){Benedettini}, {Pezzuto}, {Giannini},
  {Lorenzetti}, \& {Nisini}}]{benedettini01}
{Benedettini}, M., {Pezzuto}, S., {Giannini}, T., {Lorenzetti}, D., \&
  {Nisini}, B. 2001, \aap, 379, 557

\bibitem[{{Berrilli} {et~al.}(1987){Berrilli}, {Lorenzetti}, {Saraceno}, \&
  {Strafella}}]{berrilli87}
{Berrilli}, F., {Lorenzetti}, D., {Saraceno}, P., \& {Strafella}, F. 1987,
  \mnras, 228, 833

\bibitem[{{Bouwman}(2001)}]{bouwmanphd}
{Bouwman}, J. 2001, PhD thesis, University of Amsterdam

\bibitem[{{Bouwman} {et~al.}(2000){Bouwman}, {de Koter}, {van den Ancker}, \&
  {Waters}}]{bouwman00}
{Bouwman}, J., {de Koter}, A., {van den Ancker}, M.~E., \& {Waters},
  L.~B.~F.~M. 2000, \aap, 360, 213

\bibitem[{{Damiani} {et~al.}(2006){Damiani}, {Micela}, \&
  {Sciortino}}]{damiani06}
{Damiani}, F., {Micela}, G., \& {Sciortino}, S. 2006, \aap, 447, 1041

\bibitem[{{de Winter} {et~al.}(2001){de Winter}, {van den Ancker}, {Maira},
  {Th{\'e}}, {Djie}, {Redondo}, {Eiroa}, \& {Molster}}]{dewinter01}
{de Winter}, D., {van den Ancker}, M.~E., {Maira}, A., {et~al.} 2001, \aap,
  380, 609

\bibitem[{{di Francesco} {et~al.}(1994){di Francesco}, {Evans}, {Harvey},
  {Mundy}, \& {Butner}}]{difrancesco94}
{di Francesco}, J., {Evans}, II, N.~J., {Harvey}, P.~M., {Mundy}, L.~G., \&
  {Butner}, H.~M. 1994, \apj, 432, 710

\bibitem[{{Drew} {et~al.}(1997){Drew}, {Busfield}, {Hoare}, {Murdoch}, {Nixon},
  \& {Oudmaijer}}]{drew97}
{Drew}, J.~E., {Busfield}, G., {Hoare}, M.~G., {et~al.} 1997, \mnras, 286, 538

\bibitem[{{Dullemond}(2002)}]{dullemond02}
{Dullemond}, C.~P. 2002, \aap, 395, 853

\bibitem[{{Dullemond} {et~al.}(2001){Dullemond}, {Dominik}, \&
  {Natta}}]{dullemond01}
{Dullemond}, C.~P., {Dominik}, C., \& {Natta}, A. 2001, \apj, 560, 957

\bibitem[{{Eisner} {et~al.}(2003){Eisner}, {Lane}, {Akeson}, {Hillenbrand}, \&
  {Sargent}}]{eisner03}
{Eisner}, J.~A., {Lane}, B.~F., {Akeson}, R.~L., {Hillenbrand}, L.~A., \&
  {Sargent}, A.~I. 2003, \apj, 588, 360

\bibitem[{{Ekstr{\"o}m} {et~al.}(2008){Ekstr{\"o}m}, {Meynet}, {Maeder}, \&
  {Barblan}}]{ekstrom08}
{Ekstr{\"o}m}, S., {Meynet}, G., {Maeder}, A., \& {Barblan}, F. 2008, \aap,
  478, 467

\bibitem[{{Han} {et~al.}(1998){Han}, {Mao}, {Lu}, {Wu}, {Sun}, {Wang}, {Pei},
  {Fan}, {Tang}, \& {Ji}}]{han98}
{Han}, F., {Mao}, R.~Q., {Lu}, J., {et~al.} 1998, \aaps, 127, 181

\bibitem[{{Henning} {et~al.}(1998){Henning}, {Burkert}, {Launhardt}, {Leinert},
  \& {Stecklum}}]{henning98}
{Henning}, T., {Burkert}, A., {Launhardt}, R., {Leinert}, C., \& {Stecklum}, B.
  1998, \aap, 336, 565

\bibitem[{{Lagage} {et~al.}(2004){Lagage}, {Pel}, {Authier}, {Belorgey},
  {Claret}, {Doucet}, {Dubreuil}, {Durand}, {Elswijk}, {Girardot}, {K{\"a}ufl},
  {Kroes}, {Lortholary}, {Lussignol}, {Marchesi}, {Pantin}, {Peletier},
  {Pirard}, {Pragt}, {Rio}, {Schoenmaker}, {Siebenmorgen}, {Silber}, {Smette},
  {Sterzik}, \& {Veyssiere}}]{lagage04}
{Lagage}, P.~O., {Pel}, J.~W., {Authier}, M., {et~al.} 2004, The Messenger,
  117, 12

\bibitem[{{Leinert} {et~al.}(2003){Leinert}, {Graser}, {Przygodda}, {Waters},
  {Perrin}, {Jaffe}, {Lopez}, {Bakker}, {B{\"o}hm}, {Chesneau}, {Cotton},
  {Damstra}, {de Jong}, {Glazenborg-Kluttig}, {Grimm}, {Hanenburg}, {Laun},
  {Lenzen}, {Ligori}, {Mathar}, {Meisner}, {Morel}, {Morr}, {Neumann}, {Pel},
  {Schuller}, {Rohloff}, {Stecklum}, {Storz}, {von der L{\"u}he}, \&
  {Wagner}}]{leinert03}
{Leinert}, C., {Graser}, U., {Przygodda}, F., {et~al.} 2003, \apss, 286, 73

\bibitem[{{MacFadyen} \& {Woosley}(1999)}]{macfadyen99}
{MacFadyen}, A.~I. \& {Woosley}, S.~E. 1999, \apj, 524, 262

\bibitem[{{Maeder}(1987)}]{maeder87}
{Maeder}, A. 1987, \aap, 178, 159

\bibitem[{{Maeder} \& {Meynet}(2000)}]{maeder00}
{Maeder}, A. \& {Meynet}, G. 2000, \araa, 38, 143

\bibitem[{{Malbet} {et~al.}(2007){Malbet}, {Benisty}, {de Wit}, {Kraus},
  {Meilland}, {Millour}, {Tatulli}, {Berger}, {Chesneau}, {Hofmann}, {Isella},
  {Natta}, {Petrov}, {Preibisch}, {Stee}, {Testi}, {Weigelt}, {Antonelli},
  {Beckmann}, {Bresson}, {Chelli}, {Dugu{\'e}}, {Duvert}, {Gennari},
  {Gl{\"u}ck}, {Kern}, {Lagarde}, {Le Coarer}, {Lisi}, {Perraut}, {Puget},
  {Rantakyr{\"o}}, {Robbe-Dubois}, {Roussel}, {Zins}, {Accardo}, {Acke},
  {Agabi}, {Altariba}, {Arezki}, {Aristidi}, {Baffa}, {Behrend}, {Bl{\"o}cker},
  {Bonhomme}, {Busoni}, {Cassaing}, {Clausse}, {Colin}, {Connot},
  {Delboulb{\'e}}, {Domiciano de Souza}, {Driebe}, {Feautrier}, {Ferruzzi},
  {Forveille}, {Fossat}, {Foy}, {Fraix-Burnet}, {Gallardo}, {Giani}, {Gil},
  {Glentzlin}, {Heiden}, {Heininger}, {Hernandez Utrera}, {Kamm}, {Kiekebusch},
  {Le Contel}, {Le Contel}, {Lesourd}, {Lopez}, {Lopez}, {Magnard}, {Marconi},
  {Mars}, {Martinot-Lagarde}, {Mathias}, {M{\`e}ge}, {Monin}, {Mouillet},
  {Mourard}, {Nussbaum}, {Ohnaka}, {Pacheco}, {Perrier}, {Rabbia}, {Rebattu},
  {Reynaud}, {Richichi}, {Robini}, {Sacchettini}, {Schertl}, {Sch{\"o}ller},
  {Solscheid}, {Spang}, {Stefanini}, {Tallon}, {Tallon-Bosc}, {Tasso},
  {Vakili}, {von der L{\"u}he}, {Valtier}, {Vannier}, \& {Ventura}}]{malbet07}
{Malbet}, F., {Benisty}, M., {de Wit}, W.-J., {et~al.} 2007, \aap, 464, 43

\bibitem[{{Mannings}(1994)}]{mannings94a}
{Mannings}, V. 1994, \mnras, 271, 587

\bibitem[{{Manoj} {et~al.}(2007){Manoj}, {Ho}, {Ohashi}, {Zhang}, {Hasegawa},
  {Chen}, {Bhatt}, \& {Ashok}}]{manoj07}
{Manoj}, P., {Ho}, P.~T.~P., {Ohashi}, N., {et~al.} 2007, \apjl, 667, L187

\bibitem[{{Meynet} \& {Maeder}(2000)}]{meynet00}
{Meynet}, G. \& {Maeder}, A. 2000, \aap, 361, 101

\bibitem[{{Millan-Gabet} {et~al.}(2001){Millan-Gabet}, {Schloerb}, \&
  {Traub}}]{millan-gabet01}
{Millan-Gabet}, R., {Schloerb}, F.~P., \& {Traub}, W.~A. 2001, \apj, 546, 358

\bibitem[{{Monnier} {et~al.}(2006){Monnier}, {Berger}, {Millan-Gabet}, {Traub},
  {Schloerb}, {Pedretti}, {Benisty}, {Carleton}, {Haguenauer}, {Kern},
  {Labeye}, {Lacasse}, {Malbet}, {Perraut}, {Pearlman}, \& {Zhao}}]{monnier06}
{Monnier}, J.~D., {Berger}, J.-P., {Millan-Gabet}, R., {et~al.} 2006, \apj,
  647, 444

\bibitem[{{Monnier} {et~al.}(2005){Monnier}, {Millan-Gabet}, {Billmeier},
  {Akeson}, {Wallace}, {Berger}, {Calvet}, {D'Alessio}, {Danchi}, {Hartmann},
  {Hillenbrand}, {Kuchner}, {Rajagopal}, {Traub}, {Tuthill}, {Boden}, {Booth},
  {Colavita}, {Gathright}, {Hrynevych}, {Le Mignant}, {Ligon}, {Neyman},
  {Swain}, {Thompson}, {Vasisht}, {Wizinowich}, {Beichman}, {Beletic},
  {Creech-Eakman}, {Koresko}, {Sargent}, {Shao}, \& {van Belle}}]{monnier05}
{Monnier}, J.~D., {Millan-Gabet}, R., {Billmeier}, R., {et~al.} 2005, \apj,
  624, 832

\bibitem[{{Oudmaijer} \& {Drew}(1999)}]{oudmaijer99}
{Oudmaijer}, R.~D. \& {Drew}, J.~E. 1999, \mnras, 305, 166

\bibitem[{{Palla} \& {Prusti}(1993)}]{palla93}
{Palla}, F. \& {Prusti}, T. 1993, \aap, 272, 249

\bibitem[{{Petrov} {et~al.}(2007){Petrov}, {Malbet}, {Weigelt}, {Antonelli},
  {Beckmann}, {Bresson}, {Chelli}, {Dugu{\'e}}, {Duvert}, {Gennari},
  {Gl{\"u}ck}, {Kern}, {Lagarde}, {Le Coarer}, {Lisi}, {Millour}, {Perraut},
  {Puget}, {Rantakyr{\"o}}, {Robbe-Dubois}, {Roussel}, {Salinari}, {Tatulli},
  {Zins}, {Accardo}, {Acke}, {Agabi}, {Altariba}, {Arezki}, {Aristidi},
  {Baffa}, {Behrend}, {Bl{\"o}cker}, {Bonhomme}, {Busoni}, {Cassaing},
  {Clausse}, {Colin}, {Connot}, {Delboulb{\'e}}, {Domiciano de Souza},
  {Driebe}, {Feautrier}, {Ferruzzi}, {Forveille}, {Fossat}, {Foy},
  {Fraix-Burnet}, {Gallardo}, {Giani}, {Gil}, {Glentzlin}, {Heiden},
  {Heininger}, {Hernandez Utrera}, {Hofmann}, {Kamm}, {Kiekebusch}, {Kraus},
  {Le Contel}, {Le Contel}, {Lesourd}, {Lopez}, {Lopez}, {Magnard}, {Marconi},
  {Mars}, {Martinot-Lagarde}, {Mathias}, {M{\`e}ge}, {Monin}, {Mouillet},
  {Mourard}, {Nussbaum}, {Ohnaka}, {Pacheco}, {Perrier}, {Rabbia}, {Rebattu},
  {Reynaud}, {Richichi}, {Robini}, {Sacchettini}, {Schertl}, {Sch{\"o}ller},
  {Solscheid}, {Spang}, {Stee}, {Stefanini}, {Tallon}, {Tallon-Bosc}, {Tasso},
  {Testi}, {Vakili}, {von der L{\"u}he}, {Valtier}, {Vannier}, \&
  {Ventura}}]{petrov07}
{Petrov}, R.~G., {Malbet}, F., {Weigelt}, G., {et~al.} 2007, \aap, 464, 1

\bibitem[{{Porter} \& {Rivinius}(2003)}]{porter03}
{Porter}, J.~M. \& {Rivinius}, T. 2003, \pasp, 115, 1153

\bibitem[{{Savage} \& {Mathis}(1979)}]{savage79}
{Savage}, B.~D. \& {Mathis}, J.~S. 1979, \araa, 17, 73

\bibitem[{{Shu} {et~al.}(1994){Shu}, {Najita}, {Ostriker}, {Wilkin}, {Ruden},
  \& {Lizano}}]{shu94}
{Shu}, F., {Najita}, J., {Ostriker}, E., {et~al.} 1994, \apj, 429, 781

\bibitem[{{Skinner} {et~al.}(1993){Skinner}, {Brown}, \& {Stewart}}]{skinner93}
{Skinner}, S.~L., {Brown}, A., \& {Stewart}, R.~T. 1993, \apjs, 87, 217

\bibitem[{{Steenman} \& {Th{\'e}}(1989)}]{steenman89}
{Steenman}, H. \& {Th{\'e}}, P.~S. 1989, \apss, 159, 189

\bibitem[{{Steenman} \& {Th{\'e}}(1991)}]{steenman91}
{Steenman}, H. \& {Th{\'e}}, P.~S. 1991, \apss, 184, 9

\bibitem[{{Tatulli} {et~al.}(2007){Tatulli}, {Millour}, {Chelli}, {Duvert},
  {Acke}, {Hernandez Utrera}, {Hofmann}, {Kraus}, {Malbet}, {M{\`e}ge},
  {Petrov}, {Vannier}, {Zins}, {Antonelli}, {Beckmann}, {Bresson}, {Dugu{\'e}},
  {Gennari}, {Gl{\"u}ck}, {Kern}, {Lagarde}, {Le Coarer}, {Lisi}, {Perraut},
  {Puget}, {Rantakyr{\"o}}, {Robbe-Dubois}, {Roussel}, {Weigelt}, {Accardo},
  {Agabi}, {Altariba}, {Arezki}, {Aristidi}, {Baffa}, {Behrend}, {Bl{\"o}cker},
  {Bonhomme}, {Busoni}, {Cassaing}, {Clausse}, {Colin}, {Connot},
  {Delboulb{\'e}}, {Domiciano de Souza}, {Driebe}, {Feautrier}, {Ferruzzi},
  {Forveille}, {Fossat}, {Foy}, {Fraix-Burnet}, {Gallardo}, {Giani}, {Gil},
  {Glentzlin}, {Heiden}, {Heininger}, {Kamm}, {Kiekebusch}, {Le Contel}, {Le
  Contel}, {Lesourd}, {Lopez}, {Lopez}, {Magnard}, {Marconi}, {Mars},
  {Martinot-Lagarde}, {Mathias}, {Monin}, {Mouillet}, {Mourard}, {Nussbaum},
  {Ohnaka}, {Pacheco}, {Perrier}, {Rabbia}, {Rebattu}, {Reynaud}, {Richichi},
  {Robini}, {Sacchettini}, {Schertl}, {Sch{\"o}ller}, {Solscheid}, {Spang},
  {Stee}, {Stefanini}, {Tallon}, {Tallon-Bosc}, {Tasso}, {Testi}, {Vakili},
  {von der L{\"u}he}, {Valtier}, \& {Ventura}}]{tatulli07}
{Tatulli}, E., {Millour}, F., {Chelli}, A., {et~al.} 2007, \aap, 464, 29

\bibitem[{{Terada} {et~al.}(2001){Terada}, {Imanishi}, {Goto}, \&
  {Maihara}}]{terada01}
{Terada}, H., {Imanishi}, M., {Goto}, M., \& {Maihara}, T. 2001, \aap, 377, 994

\bibitem[{{van Boekel}(2004)}]{vanboekelphd}
{van Boekel}, R. 2004, PhD thesis, University of Amsterdam

\bibitem[{{Vink} {et~al.}(2002){Vink}, {Drew}, {Harries}, \&
  {Oudmaijer}}]{vink02}
{Vink}, J.~S., {Drew}, J.~E., {Harries}, T.~J., \& {Oudmaijer}, R.~D. 2002,
  \mnras, 337, 356

\bibitem[{{Vink} {et~al.}(2005){Vink}, {O'Neill}, {Els}, \& {Drew}}]{vink05b}
{Vink}, J.~S., {O'Neill}, P.~M., {Els}, S.~G., \& {Drew}, J.~E. 2005, \aap,
  438, L21

\bibitem[{{Vinkovi{\'c}} {et~al.}(2006){Vinkovi{\'c}}, {Ivezi{\'c}},
  {Jurki{\'c}}, \& {Elitzur}}]{vinkovic06}
{Vinkovi{\'c}}, D., {Ivezi{\'c}}, {\v Z}., {Jurki{\'c}}, T., \& {Elitzur}, M.
  2006, \apj, 636, 348

\bibitem[{{Wolf} \& {Podsiadlowski}(2007)}]{wolf07}
{Wolf}, C. \& {Podsiadlowski}, P. 2007, \mnras, 375, 1049

\bibitem[{{Woosley} \& {Heger}(2006)}]{woosley06}
{Woosley}, S.~E. \& {Heger}, A. 2006, \apj, 637, 914

\end{thebibliography}

\end{document}